\documentclass{emulateapj}

\usepackage{graphicx} 

\newcommand{\lsim}{\ \raise -2.truept\hbox{\rlap{\hbox{$\sim$}}\raise 5.truept\hbox{$<$}\ }}
\newcommand{\gsim}{\ \raise -2.truept\hbox{\rlap{\hbox{$\sim$}}\raise 5.truept\hbox{$>$}\ }}

\lefthead{Joint Analysis of LMC star clusters} \righthead{} \shortauthors{Raimondo G.}

\begin{document}

\title{Joint Analysis of near-infrared properties and surface brightness fluctuations of LMC star clusters}

\author{
G. Raimondo\altaffilmark{1} } \altaffiltext{1}{INAF-Osservatorio Astronomico di Teramo, Via M. Maggini s.n.c., I-64100 Teramo, Italy;
raimondo@oa-teramo.inaf.it }

\begin{abstract}

Surface brightness fluctuations have been proved to be a very powerful technique to determine the distance and characterize the stellar content in
extragalactic systems. Nevertheless, before facing the problem of stellar content in distant galaxies, we need to calibrate the method onto nearby
well-known systems. In this paper we analyze the properties at $J$ and $K_s$ bands of a sample of 19 star clusters in the Large Magellanic Cloud
(LMC), for which accurate near-infrared (NIR) resolved star photometry, and integrated photometry are available. For the same sample, we derive the
SBF measurements in $J$ and $K_s$-bands. We use the multi-purpose stellar population code \emph{SPoT (Stellar POpulations Tools)} to simulate the
color-magnitude diagram, stellar counts, integrated magnitudes, colors, and surface brightness fluctuations of each cluster. The present procedure
allows us to estimate the age and metallicity of the clusters in a consistent way, and provides a new calibration of the empirical $s$-parameter. We
take advantage of the high sensitivity of NIR surface brightness fluctuations to thermally pulsing asymptotic (TP-AGB) stars to test different
mass-loss rates affecting the evolution of such stars. We argue that NIR-SBFs can contribute to the disentangling of the observable properties of
TP-AGB stars, especially in galaxies, where a large number of these stars are present.

\end{abstract}

\section{Introduction}

In the study of galaxies, the age and chemical composition of the stellar components are perhaps two of the major quantities to determine. Only then,
we can trace the star formation, the chemical enrichment and the assembly history of galaxies \citep[see, e.g.][]{Renzini06}. Since first
applications, surface brightness fluctuations (SBFs) have been recognized to be effective to disentangle the evolutionary status of unresolved
stellar populations in extragalactic systems \citep[e.g.][]{Tonry+90, Blakeslee+01, Liu+02, Cantiello+03, Jensen+03, Cantiello+05}. SBFs are indeed
much more sensitive to the brightest stars in the population in a given passband than integrated luminosities \citep[e.g.][]{Liu+00}, since they are
defined as the ratio of the second to the first moment of the stellar luminosity function \citep{Tonry&Schneider88}. Thus, for instance near-infrared
(NIR) SBFs may be efficiently used to detect the presence of intermediate-age stellar populations whose light is dominated by asymptotic giant branch
(AGB) stars \citep[e.g.][]{Frogel+90}. Hence, it is now available an additional tool to supply information on stellar systems other than the
classical age/metallicity indicators based on integrated light, as broadband colors, and spectral features
\citep[e.g.][]{Worthey93,Bressan+94,Maraston98}.

Several works have been conducted to investigate the influence of AGB stars on integrated colors and spectral features in both metal-rich and
metal-poor stellar populations \citep[e.g.][]{Girardi&Bertelli98,Maraston98, Brocato+99a, Mouhcine&Lancon02, Bruzual&Charlot03, Maraston05,
Fagiolini+07}, while only a few have been devoted to SBFs. Attempts to understand how SBFs depend on the evolution parameters of AGB stars have been
made, in the optical and NIR bands, by \citet{Liu+00}, \citet{Cantiello+03}, and \citet{Raimondo+05a} following different approaches. Liu and
coworkers explored the effects of changing, by an arbitrary factor, the lifetimes of post-Main Sequence (MS) stars in stellar populations older than
1 Gyr, whereas \citet{Cantiello+03} and \citet{Raimondo+05a} linked the lifetime of AGB and thermally pulsing AGB (TP-AGB) stars, in the age range
$0.1 \lsim t (Gyr) \leq 14$, to mass-loss process, the poor knowledge of which is among major uncertainties in modeling AGB stars. These stars may
expel material at rates up to 10$^{-4}$ M$_\sun $yr$^{-1}$, eventually ejecting between 20 and 80 per cent of their initial MS mass. Thus, their
lifetime and the efficiency of the third dredge-up (TDU), which drives the formation of AGB Carbon-rich (C-type) stars, may be drastically reduced.
\citet{Raimondo+05a} found that number variations of TP-AGB stars marginally affect optical SBF magnitudes for populations older than $\sim $ 1 Gyr,
and, in general, less than $\sim $0.5 mag in the NIR, confirming the reliability of the SBF method to measure distances to spheroids
\citep{Tonry+01}. In the case of young/intermediate-age populations, instead, SBFs appear to be highly dependent on the adopted mass-loss scenario,
i.e. the number of TP-AGB stars \citep{Raimondo+05a}. An important caveat must be then understood when using SBFs to measure distances to galaxies
suspected to host intermediate-age populations.

The high sensitivity of NIR-SBFs to the evolutionary properties of AGB stars might be used, in turn, to test the prescriptions adopted in stellar
models to describe the physical processes at work in such stars. Big efforts to model stars beyond the core-helium burning phase have been recently
conducted by several authors \citep[e.g.][]{Kitsikis&Weiss07,Marigo&Girardi07,Cristallo+08,Herwig08}, however, so far, a complete understanding of
all aspects of the physics and numerical methods which properly describe AGB evolution is missing. As a matter of fact, the complex interplay between
pulsation and mass-loss process, their dependence on metallicity and stellar mass, and interior nucleosynthesis coupled with envelope chemical
enrichment are the major sources of uncertainty in predicting broad-band colors, abundances and luminosity of intermediate-age stellar populations.
Therefore, whatever the technique to derive the evolutionary status of the population is, consistency checks between model predictions and
observations for nearby resolved populations are still a prime step.

In a previous paper \citep{Raimondo+05a} we measured optical-SBF amplitudes of 11 clusters in the Large Magellanic Cloud (LMC) using data from the
Wide-Field Planetary Camera 2 (WFPC2) onboard \emph{Hubble Space Telescope (HST)}. In that work, we showed that the SBF data/model comparison suffers
of a large uncertainty when a cluster contains a small number of stars evolved off the MS, especially giant stars. Thus, we suggested that a
different algorithm has to be used to predict SBF amplitudes for star clusters, due to the paucity of stars in such systems. \citet{Gonzalez+04} used
a different approach: they built up 8 super-clusters by coadding clusters in the LMC and the Small Magellanic Cloud (SMC) having the same
\citet[][SWB]{Searle+80} class. The authors measured NIR-SBFs of these super-clusters using the Second Incremental and All Sky Data releases of the
Two-Micron All Sky Survey (2MASS). To some extent, grouping clusters reduces stochastic effects due to small numbers of stars on fast evolutionary
phases, however, this procedure suffers of the uncertainty in defining the membership of the cluster to the SWB classes as well as of the uncertainty
of the relationship between the \emph{s}-parameter and the cluster age. Moreover, \citet{Mouhcine+05} and \citet{Raimondo+05a} argued that the
discrepancy between predicted and observed SBFs of some super-clusters is likely due to observational reasons, as contamination by foreground
sources, and low-quality photometry of stars in the central, most crowded, regions. Therefore, obtaining new SBF measures from NIR data having better
spatial resolution and sky stability is a necessary step toward constraining more tightly stellar population models.

In this paper we analyze the properties of a sample of nearby populous intermediate-age stellar clusters expected to host TP-AGB stars. We study 19
star clusters of LMC observed in the NIR by \citet{Ferraro+04} and \citet{Mucciarelli+06}, and analyze their color-magnitude diagram (CMD),
luminosity functions (LFs), integrated colors, magnitudes, and SBF amplitudes. From resolved-star photometry and total fluxes we measure SBFs in the
$J$ and $K_s$ pass-bands. The multi-purpose stellar population synthesis code \emph{SPoT} (Stellar Population
Tools\footnote{http://www.oa-teramo.inaf.it/SPoT}, \citealt{Raimondo+05a}) is used to derive theoretical SBF amplitudes, CMDs, integrated magnitudes,
and colors of each cluster. We explore the sensitivity of SBFs to TP-AGB stars by changing the mass-loss rates. In the models presented here, the
mass-loss rate is assumed to be the main parameter triggering the lifetime of TP-AGB stars, whereas the other relevant physical quantities quoted
above are assumed to be similar in each models. The goal of the present paper is then twofold. First, we intend to reproduce the cluster's CMD, LFs,
colors, and SBFs by means of a unique theoretical framework. Second, we address whether a connection between SBFs and TP-AGB star properties (number)
and cluster age can be discerned in SBF data. The paper is organized as follows. \S~\ref{s:models} details ingredients of the stellar population
synthesis (SPS) code, including a brief description of the method adopted to derive integrated quantities (colors and SBFs). In \S~\ref{s:data}, we
present the cluster sample and data, and in \S~\ref{s:datamodels} we compare synthetic CMDs, LFs, integrated colors, magnitudes and SBFs to the
corresponding observed quantities. The final section provides a summary and discussion of our results.

\section{Stellar Population Synthesis Code}
\label{s:models}

The SPS models presented in this paper are based on an updated version of the stellar population synthesis code \emph{SPoT}, designed to reproduce
the properties of both resolved and unresolved stellar populations. SPS model predictions were applied and tested on star clusters of the Galaxy and
MCs in previous papers \citep[e.g.][]{Brocato+99a,Brocato+00,Brocato+03,Raimondo+02,Raimondo+05a,Cantiello+03,Cantiello+07b}. In this section, we
briefly describe the main ingredients, techniques to derive integrated quantities, and outline changes. We refer the reader to the cited papers for
more details.

Differently from most SPS codes that are based on isochrones, the \emph{SPoT} code directly relies on stellar evolutionary tracks. The present
version utilizes the updated evolution stellar library by \citet{Pietrinferni+04}\footnote{The stellar evolutionary library by
\citet{Pietrinferni+04} (BaSTI web-site: http://www.oa-teramo.inaf.it/BaSTi) have been recently recomputed by the authors.} for masses $0.5 \leq
M/M_{\odot} \leq 11 $. The stellar models cover all the evolutionary phases from MS up to carbon ignition or the onset of thermal pulses (TPs). This
allows us to analyze stellar populations in the age range from $\approx 50$ Myr to 14 Gyr. Stars of mass $0.1 \lsim M/M_{\odot} < 0.5 $ are from
\citet{Brocato+98}. Masses lower than this limit (central H-ignition, $M \approx 0.08 \, M_{\odot}$) do not significantly contribute to the total
mass of the cluster \citep[see e.g.][]{Chabrier&Mera97}, and so they are not taken into account.

The mass of each star is randomly generated by using Monte Carlo techniques, while the mass distribution is shaped by the Initial Mass Function (IMF)
from \citet{Kroupa01}. The evolutionary line of each mass is then calculated by interpolating the available tracks in the mass grid.

\subsection{Horizontal Branch, AGB and TP-AGB stars} \label{ss:hb}

The code is suited to simulate the color distribution of He-burning stars on the horizontal branch (HB) as a function of age, metallicity and
mass-loss rate in old stellar populations \citep[$t \gsim 5$ Gyr; e.g.][]{Brocato+99b,Raimondo+02}. The mass loss suffered by RGB stars is evaluated
according to the Reimers formulation, where the Reimers' parameter \emph{$\eta^R_i$} ($i$ refers to the $i$-th RGB star) follows a Gaussian
distribution-function with a mean value $\langle {\eta^R} \rangle = 0.4$ in our \emph{standard models}. Thus, a generic RGB star losses mass at rate:

\begin{equation}
\dot M^{R}_i = - 4\cdot 10^{-13}\eta^R_i \cdot L_i R_i/M_i, \label{eq:reimers}
\end{equation}

\noindent where $L_i, R_i, M_i$ are respectively the star luminosity, radius and total mass in solar units. The result of this procedure is that
He-burning stars are spread on the HB as observed in Galactic globular clusters \citep[e.g.][]{Brocato+00,Raimondo+02}. SBF amplitudes of stellar
populations older than $\sim  $ 5 Gyr were linked to the HB morphology as a function of $\eta^R$ in \citet{Cantiello+03}.

The evolution of stars off the zero-age HB till the double-shell phase is followed by interpolating the evolutionary tracks of
\citet{Pietrinferni+04}. In the case of intermediate-age populations, the stellar mass evolving off the turn-off (TO) point is high enough to prevent
the RGB phase (i.e. no degenerate He core is developed), so that mass loss does not affect sizeably the color distribution of He-burning stars.

Beyond the early-AGB phase, stars with a MS mass in the range $M \sim 1-8$ M$_\sun$ (the upper limit, called $M_{up}$, depends on metallicity)
undergo the TP-AGB phase. TP-AGB synthetic models \citep[e.g.][]{Renzini&Voli81, Iben&Renzini83} and hybrid models, which combine aspects of
synthetic and full evolution \citep[e.g.][]{Marigo98}, were developed with the aim to supply simplified descriptions of stellar evolution in this
phase, using recipes and descriptions based on full evolutionary models and/or derived from empirical calibrations. In our code a specific routine
evaluates the properties of expected TP-AGB stars by integrating the analytic formulas of \citet[][hereandafter WG98]{Wagenhuber&Groenewegen98}, as
described in \citet[][see Appendix A]{Raimondo+05a}. It calculates the time evolution of the stellar core mass and luminosity, the starting point
being the helium-core mass, luminosity, and temperature at the first thermal pulse. The WG98 description includes three important effects: ($i$) the
first pulses do not reach the full amplitude; ($ii$) the hot bottom burning process that occurs in massive stars; and ($iii$) the TDU, even though
this last phenomenon is treated in a simple way. In fact, in their models WG98 assumed that there is dredge-up only if the core mass is higher than a
critical value and that the free parameter $\lambda$, describing the dredge-up efficiency, is constant. The limit of this formulation is that these
descriptions and formulae do not account for the influence of abundance changes and nucleosynthesis in general on the evolution of the star. In the
present work, the effective temperature ($T_{eff}$) is evaluated using relations of \citet[][hereafter W90]{Wood90}, slightly shifted ($\Delta \log
T_{eff} = -0.05$) to match present evolutionary tracks.

All stars are oxygen-rich (O-rich) when they enter the AGB phase. Whether or not they become C-rich stars primarily depends on the efficiency of the
TDU occurring in the TP-AGB phase and on the extent and time-variation of the mass loss \citep[e.g.\ ][]{Straniero+97}. However, several effects play
a role. In low-metallicity stars ($Z < 0.004$), the amount of oxygen in the envelope is so low that a few thermal pulses are sufficient to convert an
O-rich star into a C-star \citep[][]{Renzini&Voli81}. Moreover, the lower the metallicity the lower the minimum mass for the onset of TDU
\citep[e.g.][]{Straniero+03}. On the other hand, strong mass-loss episodes in TP-AGB stars may delay or even prevent the TDU occurrence and the
formation of C-rich low-mass stars \citep{Marigo+99}. To introduce the metallicity dependence of the O- to C-star ratio, we consider the relation
empirically derived in galaxies of the Local Group by \citet{Groenewegen06}. So that, we assume that the ratio of the duration of the two sub-phases
for each star with a given metallicity is constant.

Finally, post-AGB evolution experienced by stars before entering the white dwarf cooling sequence has a negligible impact in simulating the NIR
properties of individual clusters, thus in the present work, such kind of stars are not considered.

\subsection{Mass loss of TP-AGB stars} \label{ss:TPmassloss}

Major uncertainties in modeling the TP-AGB phase are related to the duration of this phase and to variations of stellar luminosity and chemistry. As
already stated, the duration is highly triggered by various physical mechanisms, and primarily driven by the efficiency of mass-loss processes. To
account for current uncertainties in the mass-loss rate determinations, we consider four different mass-loss prescriptions, and assume that the other
relevant physical quantities are described by the WG98 and W90 formulations. Our aim is to investigate \emph{if and how} they produce sizable effects
on the foreseen observational quantities, SBFs in particular. Our final goal is to test the sensitivity of SBFs to the number of TP-AGB stars, and to
investigate whether SBF predictions/data of well studied MCs star clusters may give suggestions about the number of TP-AGB stars.

In \citet{Raimondo+05a} we adopted the mass-loss rate formulation of \cite{Bloecker95} based on the dynamical theoretical investigation on the
atmospheres of Mira-like variables by \citet{Bowen88} and valid for long period variable (LPV) stars with periods $P>100$ days:

\begin{equation}
\dot M^{B95} = 4.83 \cdot 10^{-9} \eta^{B} M^{-2.1} L^{2.7} \dot M^{R} \label{eq:b1}
\end{equation}

\noindent where the index $i$ is suppressed for simplicity; $M$ is the MS mass, $\eta^{B}$ is a scaling factor, equal to unity in
\citet{Raimondo+05a}, and $\dot M^{R}$ is the Reimers mass-loss rate with $\eta^R=1$.

The four models considered in the present paper are defined by the following mass-loss laws:

\vspace{0.2cm}

\noindent $-$ \emph{Model A}: The mass-loss rate of \citet[][Eq. \ref{eq:b1}]{Bloecker95} with $\eta^B=0.01$ and $\eta^B=0.05$ is applied to stars
with $ M \geq 1.2 M_\sun$, whereas for lower masses the improved description for cool winds of tip-AGB stars by \citet{Wachter+02} is adopted:

\begin{equation}
\dot M^{W02} = -4.52 - 6.81 \log (T_{eff}/2600) + 2.47 \log (L/10^4L_\sun) -1.95 \log (M) \label{eq:w02}
\end{equation}

\vspace{0.3cm}

\noindent $-$ \emph{Model B}: The original Bloecker's law with $\eta^B=1$ is applied to the full range of masses.

\vspace{0.3cm}

\noindent $-$ \emph{Model C}: The empirical mass-loss rate derived by \citet{vanLoon+05} for O-rich dust-enshrouded AGB stars in the LMC is assumed.
Through the modeling of spectral energy distributions, the authors estimated a mass-loss rate as a function of stellar luminosity and effective
temperature:

\begin{equation}
\log \dot M^{V05} = -5.65 + 1.05 \log(L/10000 L_\sun) - 6.3 \log(T_{eff}/3500K)
\end{equation}

\vspace{0.3cm}

\noindent $-$ \emph{Model D}: The law proposed by \citet{vanLoon06} for very bright stars with dust-driven mass-loss is adopted

\begin{equation}
\dot M^{V06} = 1.5 \times 10^{-9} Z[Z_{\sun}]^{-0.5}\, L[L_{\sun}]^{0.75}\, A_V^{0.75} \label{eq:V06}
\end{equation}
Here, $A_V$ is the visual extinction ranging from about 0.01 up to 100 \citep{vanLoon06}.  In our synthesis models we assumed $A_V$ as a free
parameter, and find \emph{a posteriori} that values in the range 0.01$-$1 are suitable to reproduce present NIR data. This is because most
dust-enshrouded AGB stars are visible at longer wavelengths compared to the ones used here ($J$ and $K_s$ bands). Moreover, from the analysis of a
sample of oxygen-rich AGB stars, \citet{Heras+05} determined a mass-loss rate in the range $5 \times 10^{-8} \div 10^{-5}$ M$_\sun$ yr$^{-1}$ and a
visual optical depth of $0.03 \div 0.6$ for stars with a small amount of dust in their envelopes.

\begin{figure*}[t]
  \plotone{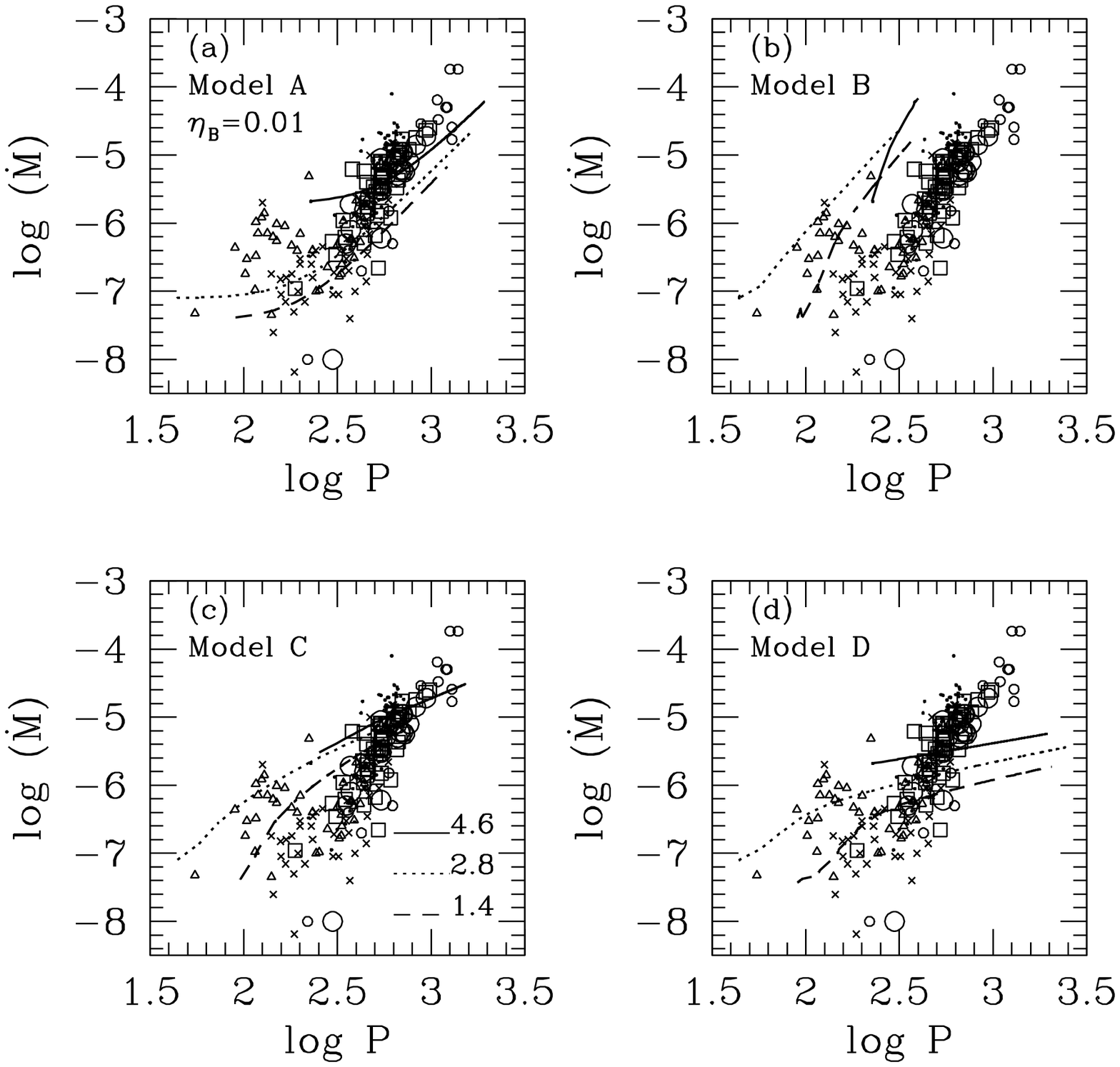}
 \caption{Observed periods and mass-loss rates of AGB stars from the literature: O-rich and C-rich
 stars in the LMC \citep[][small and big open circles, respectively]{Whitelock+03}; C-stars of LMC
 \citep[][squares]{Groenewegen+07}; Galactic C-rich stars
 from \citet[][dots]{Whitelock+06} and \citet[][crosses]{Schoier+01}; Galactic AGB stars \citep[][triangles]{Winters+03}.
 Data are compared to the theoretical sequences
 of three masses as labeled, metallicity $Z=0.008$, and mass-loss formulations from:
 \citet{Bloecker95} with $\eta^B = 0.01$ ($M\geq 1.2$ M$_\sun$) and \citet{Wachter+02} ($ M < 1.2$ M$_\sun$) in panel $(a)$;
 \citet{Bloecker95} with $\eta^B = 1$ for all masses in panel $(b)$; \citet{vanLoon+05} and \citet{vanLoon06} in
 panel $(c)$ and $(d)$, respectively.}
 \label{fig:dmp}
\end{figure*}

To illustrate differences between models, we computed the mass-loss rate evolution of three TP-AGB stars of $M=1.4$, 2.8, and 4.6\,M$_\sun$ at fixed
metallicity ($Z=0.008$). The resulting evolutionary paths as a function of the pulsation period are reported in Figure~\ref{fig:dmp}. Theoretical
fundamental periods were calculated from period-mass-luminosity relations for long period variables of \citet{Fox&Wood82}, as reported by
\citet{Marigo&Girardi07}, under the assumption that stars pulsate everywhere on the TP-AGB phase, even if an instability strip should be considered
\citep{Groenewegen&deJong94b}. Theoretical sequences are qualitatively compared with empirical mass-loss rates, derived from IR fluxes and CO radio
line emission, and pulsation periods from observed light curves of a sample of obscured C- and O-rich stars \citep{Whitelock+03} and C-stars
\citep{Groenewegen+07} in LMC. In the figure Galactic AGB stars observed by \citet{Whitelock+06}, \citet{Winters+03}, and \citet{Schoier+01} are also
plotted. As remarked by \citet{Whitelock+06}, the two populations do not show remarkably differences in color and period relations. A similar
conclusion can be inferred for the mass-loss rate-period relation, since at a given period mass-loss rates for LMC stars are in agreement with those
observed in Galactic stars \citep[see also][]{Groenewegen+07}. Despite a large spread at shorter periods, data suggest that the mass-loss rate
increases strongly with period greater than $\log P \sim 2.6$. A less steep dependence can be argue at lower periods \citep[see
also][]{Straniero+06}, even if data (Galactic stars mostly) are more spread, as they suffer of the uncertainty on the distance, which appears in the
expression used to derive mass-loss rate \citep[see e.g.][]{Whitelock+06}.

From Figure~\ref{fig:dmp} it appears that in the case of \emph{model A} (panel \emph{(a)}) the evolutionary lines nicely reproduce the observed
trend. As a matter of fact, the bulk of stars are fairly reproduced by stars having masses approximatively in the range $\approx 1.4 \div 4.5
M_{\sun}$, and mass-loss rates reaching values as high as $\approx 8 \times 10^{-5} M_{\sun}$yr$^{-1}$. A few stars with higher mass-loss rates are
O-rich stars belonging to the LMC field population (small open circles). Note that here we plot models with a single metallicity, while data refer to
galaxy populations, consisting of a mixture of stars with different age, metallicity and mass. There are no studies that conclusively show that the
mass-loss rate explicitly depends on metallicity \citep{Zijlstra+96}, however metallicity may have effect on grain growth, the number density and
size of grains, thus altering the mass-loss rates from the stars. Finally, mass-loss rates may strongly depend on stellar pulsation. Stars of high
metallicity, e.g. $Z=0.02$, and mass are able to reach lower effective temperatures and higher mass-loss rates. In our models the central star
typically has $T_{eff}$ in the range $\sim 2500-3800$ K, in fair agreement with estimations found by \citet{Groenewegen+07} and \citet{vanLoon+05}
for MCs field populations, even thought the latter sample includes cooler stars. If $\eta^B=1$ (\emph{model B}) the mass-loss process is very
efficient in stripping away the stellar envelope, and the evolution results steep and rapid. The formulation of \citet{vanLoon+05} applied to our
TP-AGB modeling produces a rapid increase of the mass-loss rate since the beginning of the phase (Figure~\ref{fig:dmp}, \emph{model C}). The relation
from \citet{vanLoon06} produces a less steep behavior and longer periods at a given mass-loss rate  (Figure~\ref{fig:dmp}, \emph{model D}). However,
we recall this formulation critically depends on $A_V$. In the rest of the paper, unless explicitly stated otherwise, we present results of models
$A$ and $C$, even though computations and comparison for models $B$ and $D$ have also been performed.

As conclusive remark, we notice that changing the mass-loss rate prescription may also affect the TP-AGB star structure, so that here we have
performed a qualitative comparison between results of our synthetic modeling of TP-AGB stars and observations. The detailed treatment of the
evolution of stars along the complex TP-AGB phase requires accurate TP-AGB evolution stellar models, for instance the modeling of the two stages (O-
and C-rich phases) as a function of stellar mass and metallicity, by taking into account the internal structure evolution, stellar pulsation,
mass-loss mechanism, and dust and grain formation \citep[see e.g.,][]{Izzard+04,Marigo&Girardi07,Cristallo+09}. Nevertheless, the agreement between
the data and the general picture that we have presented is already very encouraging to investigate how these scenarios affect SBF predictions, as we
do in the following sections.

\subsection{Integrated magnitudes}
\label{ss:imteo}

To compute the cluster total fluxes we assume that the integrated luminosity is dominated by light emitted by their stellar component. On this basis,
the total fluxes mainly depend on two quantities: \emph{1)} the flux $f_{i}$ emitted by the $i$-th star of mass $M$, age $t$, luminosity $L$,
effective temperature $T_{eff}$, and chemical composition ($Y$, $Z$):

\begin{equation}
 f_{i}[L(M,t,Y,Z),T_{eff}(M,t,Y,Z),Y,Z]
\end{equation}
and \emph{2)} the number of stars of mass $M$ in a population counting $N$ stars, $\Phi(M,N)$. $f_{i}$ is defined by the stellar evolution library
and the temperature-color transformation tables, while $\Phi(M,N)$ is strictly related to the IMF. The Monte Carlo procedure we adopted is essential
to simulate poorly populated stellar systems, like those we study in the following. It ensures that fast, under-sampled evolutionary phases, are
correctly treated from a stochastic point of view. The above quantities are combined to calculate the total integrated flux $F$ in a given
photometric band:

\begin{equation}
 F[N,t,Y,Z] = \sum _{i=1}^N f_{i}.
\end{equation}

To account for stochastic effects, we computed a large number ($N_{sim}$) of independent simulations for each set of population parameters (age,
chemical composition, cluster mass, etc.). So that, we obtained statistical distributions of magnitudes (colors), produced by stochastic variations
in the number and properties of bright and rare stars. For the purposes of this paper, $N_{sim}=200$ is appropriate to reproduce the observed
clusters (see also Figure~2 in \citealt{Raimondo+05a}).

The conversion from theoretical quantities to magnitudes and colors is based on the BaSeL library \cite[][and reference therein]{Westera+02}, with
the exception of C- and O-rich stars, whose colors are derived from spectra of \citet{Lancon&Mouhcine02}. The homogenized photometric system of
\citet{Bessell&Brett88} is adopted, so that in the following magnitudes and colors involving the $K$ filter are transformed into $K_s$ filter
according to \citet{Carpenter01}. This assumption is adequate as extensively explained in \citet{Pessev+06}.

\subsection{Surface Brightness fluctuations}
\label{ss:SBFteo}

If the type and flux of stars that comprise the stellar population luminosity function are known, the SBF magnitude is defined as
\citep{Tonry&Schneider88}:

\begin{equation}
\overline {M} = -2.5 \log \ [ \frac {\sum_{i=1}^{N} f_i^2} {\sum_{i=1}^{N} f_i} ] \label{eq:eqTS}
\end{equation}

\noindent where $\overline {M} $ relies upon the \emph{Poissonian statistics}. This is the case of our synthetic stellar population models, in which
the flux of each star is known. In the case of a large number of independent simulations, as adopted here, the average SBF magnitude can be defined
as

\begin{equation}
\overline {M} =   \frac {\sum_{j=1}^{N_{sim}} \overline {M}_j } {N_{sim}}, \label{eq:eqTS2}
\end{equation}
and the statistical uncertainty is derived as the standard deviation of the $\overline {M}_j$ distribution ($j=1, N_{sim}$).

\section{Cluster sample and data}
\label{s:data}

The NIR photometry of 19 clusters in the LMC is from \citet{Ferraro+04} and \citet{Mucciarelli+06}, obtained with the SOFI imager/spectrometer
mounted on the ESO 3.5m NTT. Each cluster typically contains 1000-1200 objects, with the exception of a sub-sample of poorly populated clusters (such
as NGC~2190, NGC~2209, NGC~2249, NGC~1651, NGC~2162, and NGC~2173) and an handful of rich clusters (such as NGC~1783, and NGC~1978) with more than
1900 stars.

The clusters are listed in Table~\ref{tab:cluster_info} together with some properties. The age estimations from \citet{Mucciarelli+06} have been
derived through the $s$-parameter calibration of \citet{Girardi+95}, based on stellar models computed with a certain amount of convective overshoot.
$[Fe/H]$ determinations from either spectroscopic or photometric data are also reported. The last three columns list the number of C-stars identified
by various authors and the  $s$-parameter from \citet{Elson&Fall85}. The membership of C-rich giant stars is not well established in some cases. For
instance, a giant star in NGC~2136/37 is identify as a C-star with $K \simeq 10.7$ mag and $J-K=1.53$ mag since the work by \citet{Aaronson&Mould85},
even its membership is not confirmed \citep{Frogel+90}.

We fixed the total $V$-magnitude of each cluster to be equal to the value published by \citet{Goudfrooij+06}, who measured integrated-light
photometry in Johnson-Cousins $V, R$ and $I$ for a sample of 28 star clusters in the MCs. From their Table~A1 we selected measurements at an aperture
radius corresponding to $1\arcmin$.5, the same area used to the CMD analysis (see below). For clusters not included in the sample of
\citet{Goudfrooij+06}, we consider measurements by \citet{VanDenBergh81}, bearing in mind that they generally refer to a smaller aperture size
($\sim$ 50-60$\arcsec$). After a rescaling, the formal difference between the two sets of measures is negligible, being of the order of $(V_G - V_V)
= -0.03$ mag \citep{Goudfrooij+06}.

The distance modulus of LMC is settled to be $(m-M)_0 = 18.40\pm0.10 $ mag, on the basis of recent estimations \citep[e.g.][]{Grocholski+07,
Testa+06, Walker+01}. It is worth mentioning that a variation of the LMC distance within the above uncertainty, e.g. the adoption of the usual
distance value $(m-M)_0 = 18.50$ mag, does not affect the results presented below. Similar consideration can be done for the Galactic extinction law,
here we use $R_V = 3.4$ \citep{Gordon+03}. The conversion of $A_V$ to extinction coefficients in other photometric bands was done using formulae in
\citet{Cardelli+89}.

\section{Data/Models comparison}
\label{s:datamodels}

The theoretical framework described in \S~\ref{s:models} was adopted to derive fully consistent synthetic CMDs and integrated quantities. The
comparison with \emph{resolved} features of well studied star clusters (\S~\ref{s:data}) provides the opportunity to calibrate NIR models in detail,
by fitting the observed CMDs, star counts and luminosities, and at the same time integrated and SBF magnitudes. We take advantage of this joint
analysis to infer indication on the efficiency of SBFs to estimate ages and metallicities of unresolved intermediate-age systems and to constrain the
number and photometric properties of bright red giant stars.

\subsection{CMD, luminosity function and stellar counts}
\label{ss:cmd}

The procedure to fit models to data can be summarized as follows. First, a series of synthetic CMDs were computed to estimate the age and chemical
composition by fitting the observed CMD, and the cumulative luminosity function (CLF)  of each cluster. The age step is 10\% of the expected age,
while the metallicity is given by the available grid of stellar evolution models, i.e. no-interpolation was done. Each synthetic CMD includes a
fraction of detached binaries whose mass ratio was distributed homogeneously between 0.7 and 1, as found in several LMC clusters (e.g. NGC~1818:
\citealt{Elson+98}; NGC~1866: \citealt{Brocato+03, Barmina+02}; NGC~2173: \citealt{Bertelli+03}). The fraction of binaries is a free parameter
ranging from 10\% up to 70\% of the total. Age and metallicity evaluations are given by simultaneously fitting the distribution (CLF) and photometric
properties (CMD) of stars in the cluster. Note that the computed cluster's $V$-magnitude ($M_V^{tot}$) is equal to the observed value. For the
handful of clusters whose $M_V^{tot}$ comes from the compilation of \citet{VanDenBergh81}, the systematic difference of the aperture-size was taken
into account. Once the best-fitting parameters (age, metallicity, and fraction of binaries) were established, we computed a series of 200 synthetic
CMDs in order to take into account stochastic effects (see \S~\ref{ss:imteo} and \S~\ref{ss:SBFteo}). It is relevant to emphasize that random
extractions of stellar masses are fully independent, even though the same set of input parameters is assumed. Moreover, when simulating the CMD we
did not draw the same number of stars as observed, instead the number of stars placed in different regions of the synthetic CMD turns out from
computations when the total magnitude is assumed to be equal to the observed one.

\begin{figure*}[t]
  \plotone{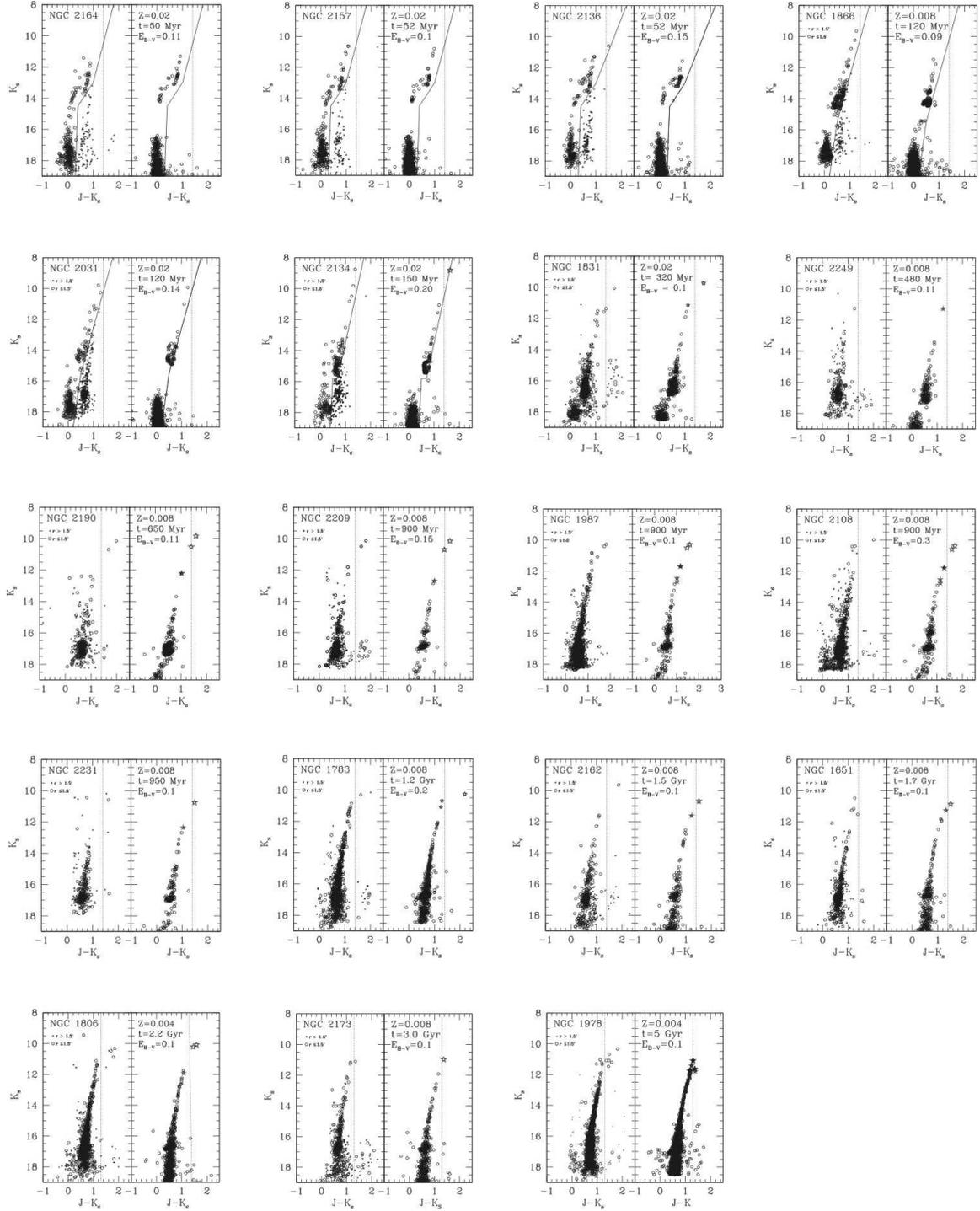}
 \caption{Left panels: Observed NIR-CMDs. Stars within 1$\arcmin$.5 are plotted as larger circles, while
  stars at $r>$1$\arcmin$.5 as dots. We plot sources which were detected in both $J$ and $K_s$ bands and have no artifacts in these bands.
  In the first six clusters the CMD region at the left (right) of the solid line is likely dominated by the cluster (field) population.
  Right panels: One of the simulated CMDs (\emph{model A}). The adopted metallicity (Z), age and $E_{B-V}$
  are labeled from top to bottom. Predicted C-rich and O-rich TP-AGB stars are indicated as stars and asterisks, respectively.
  The vertical dotted lines indicate the rough separation between O- and C-rich stars based on their $J-K_s$ color (see text).}
  \label{fig:cmd}
\end{figure*}

Figure~\ref{fig:cmd} shows observed (left) and synthetic (right) [$K_s, J-K_s$]-CMD of each cluster. The synthetic CMD plotted is one of the 200
simulations computed in the framework of \emph{model A}. C-type AGB stars are identified as stars, while M-type stars as asterisks. The vertical
dotted lines indicate the rough separation between C-rich (redder) and O-rich (bluer) stars on the base of their $J-K_s$ color: ($J-K_s) \approx 1.4$
mag at $Z\geq 0.008$ and $(J-K_s) \approx 1.3$ mag at $Z=0.004$, as empirically determined by \citet{Cioni+01, Cioni+03} in the LMC and SMC and
confirmed by the cross-correlation with spectroscopic surveys for example in SMC by \citet{Raimondo+05b}. The clusters are arranged for increasing
age, from 50 Myr to 5 Gyr. Metallicity ranges from $Z=0.02$ down to $Z=0.004$. The reddening value ($E_{B-V}$) labeled in each panel is the mean
value of the measures available in the literature.

As pointed out by \citet{Mucciarelli+06} and \citet{Ferraro+04} a significant contribution of LMC field stars is present in all clusters (see their
figures 1-4, and figures 2-4, respectively). However, in young clusters a blue sequence is clearly visible at $-0.3 \lsim (J - K_s) \lsim 0.3$ and
$K_s \gsim 15.5$, corresponding to the brightest end of the cluster MS. At $K_s \sim 14$ mag the region occupied by He-burning stars is also
recognized. So that, it is possible to confidently identify cluster's stars  as those placed at the left of the solid lines reported in
Figure~\ref{fig:cmd} (first six plots), notwithstanding a residual uncertainty remains to establish the membership of stars with $K < 12-13$ mag
(likely AGB stars). In the older clusters the observational limit ($K_s \approx 18.5$) does not permit to detect the TO point and for ages greater
than $\approx 400 $ Myr it is not possible to easily identify field stars, since they overlap the cluster population. For all these reasons, to
minimize the field contamination and to better estimate the cluster age and metallicity, the fit-procedure was applied to stars as far as $r=$
1.5$\arcmin$ off from the cluster center (large circles in the left panels of Figure~\ref{fig:cmd}), as suggested by \citet{Ferraro+04}. In addition,
we used the photometry of the control field a few arcmin away from each cluster to subtract the contribution of field stars to the observed CLF,
after to have normalized and rescaled the data to the selected cluster's area. The resulting CLFs are used in the fit-procedure.

\begin{figure*}[t]
  \plotone{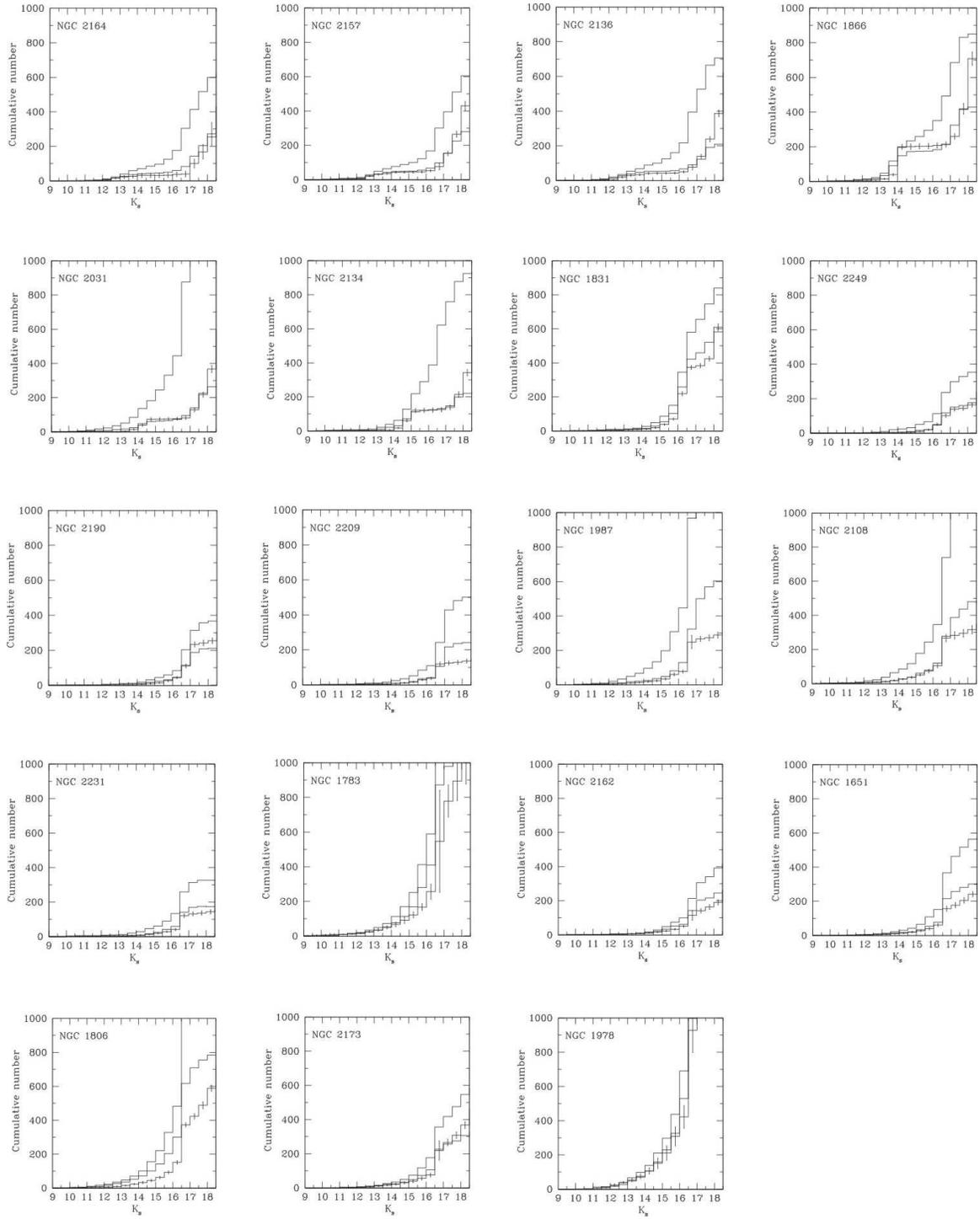}
  \caption{Cumulative $K_s$-LF (CLF) of each cluster: Dotted lines are CLFs from all observed stars, long-dashed lines
  are CLFs after the subtraction of field stars. Only the cluster's area within 1$\arcmin$.5 is considered.
  Solid lines represent the mean synthetic CLF obtained from 200 simulations in the framework of
  \emph{model A}. The estimated error is also plotted in each bin.}
  \label{fig:clf}
\end{figure*}

Figure~\ref{fig:clf} compares observed and synthetic CLFs. Synthetic CLFs well reproduce the observed counterparts, especially at $K_s \lsim 17-17.5$
mag. This is a reliable limit free by crowding effects and completeness \citep{Mucciarelli+06}. For two clusters (NGC~1806 and NGC~1783) the
agreement is limited to the brightest part of the CLF ($K_s \lsim 14-15$ mag). Recent optical observations have suggested the presence of two stellar
populations in the area of these two clusters, with an age difference of $\approx 300$ Myr \citep{Mackey+08}.  The NIR data used here do not permit
to reach the TO point, however our best age estimation for NGC~1783 is about 1.2 Gyr, in fair agreement with the value recently suggested by
\citet[][$t \sim 1.4$ Gyr]{Mucciarelli+07} and slightly younger than the interval suggested by \citet[][1.6-2.2 Gyr]{Mackey+08}, both derived from
isochrones which include some overshooting. Other few clusters are recognized to be complex systems, for instance NGC~2136 and NGC~1987. The former
is a component of a potential triple cluster system \citep{Hilker+95}. The latter is a very poorly-populated cluster lying in a highly
field-contaminated region \citep{Corsi+94, Ferraro+04}, an occurrence that appears to be confirmed in Figure~\ref{fig:clf}. Corsi and co-workers in
their optical CMD discriminated three different populations: a very old field population, a young population producing a blue plume extending the
bright MS, and the cluster population. Moreover, \citet{Frogel+90} identified two C-stars in the cluster area, whose membership, however, is
determined uncertain for one of the two.

As our models \emph{A-B-C-D} differ only for the treatment of the TP-AGB phase, there is a good agreement between observed and synthetic CMDs till
the early-AGB either for \emph{model A} or models \emph{B}, \emph{C} and \emph{D}. Thus similar ages and metallicities are predicted from all the
models. In contrast, the four models produced a different number of stars in the subsequent TP-AGB phase (see below).

Table~\ref{tab:cluster_synthetic} summarizes the age and metallicity, the mean $M_V$ and $M_{K_{s}}$ absolute magnitudes, and $J-K_{s}$ and $H-K_s$
colors (\emph{models A}) for each cluster. As we used the available metallicity grid without making any interpolation, one can consider as
metallicity uncertainty the difference between the best-fit value and the nearest two values in the grid.

\begin{figure*}[t]
  \plotone{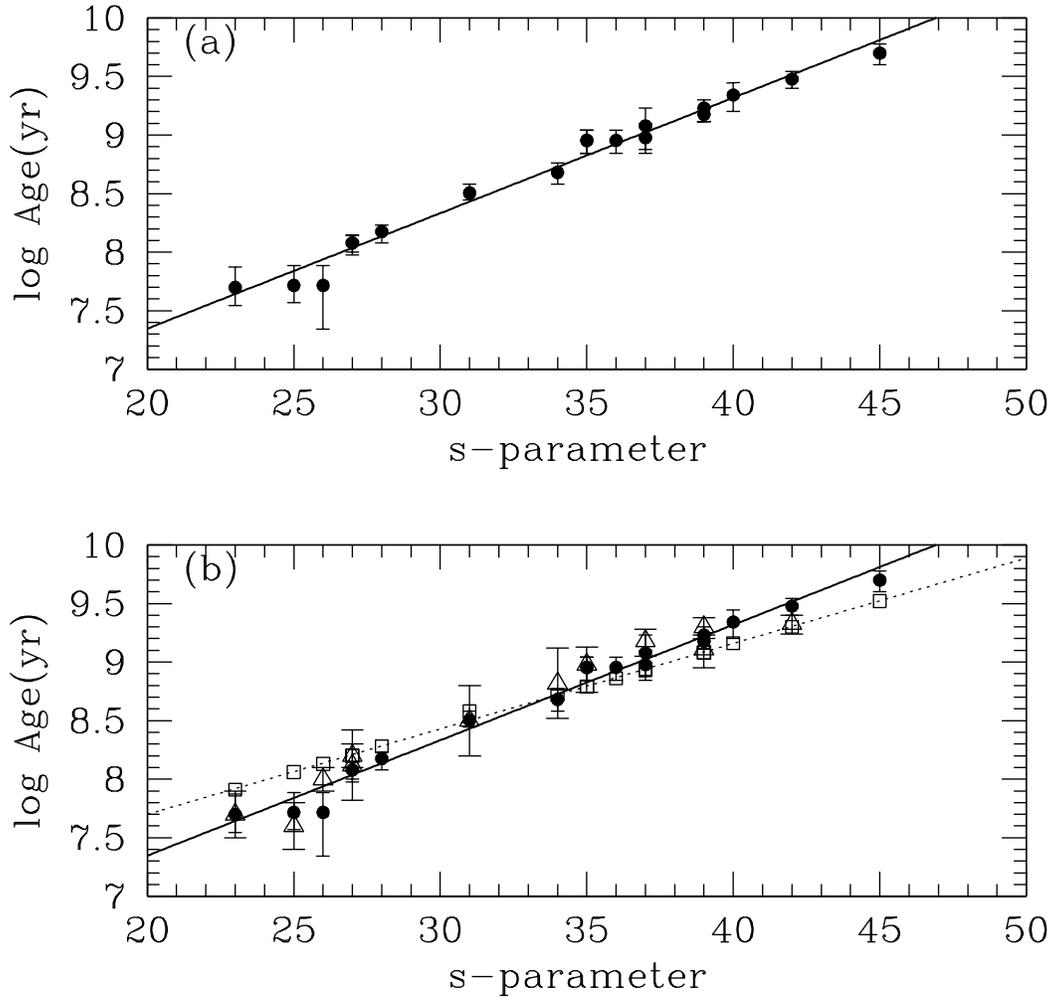}
  \caption{Panel (a): Age estimations (filled circles) and age-$s$
  calibration (solid line) from the present analysis.
  The $s$ parameter comes from \citet{Elson&Fall85}. Panel (b): The present age-$s$
  calibration (solid line) is compared to ages from (\citealt{Mucciarelli+06}, open squares, dotted line),
  and ages estimated from the CMD by various authors (see Col.2 in Table~\ref{tab:cluster_info}, open triangles).}
  \label{fig:s-par}
\end{figure*}

From the age estimations in Table~\ref{tab:cluster_synthetic} and $s$ values from Table~\ref{tab:cluster_info} we obtain the following age-$s$
parameter calibration (Figure~ \ref{fig:s-par}$a$):

\begin{equation}
\log t = (5.37\pm0.12) + (0.099 \pm 0.003 ) \times s \label{eq:spar}
\end{equation}
Our calibration based on canonical stellar evolutionary tracks is compared to that of \citet{Girardi+95} and ages from the literature in
Figure~\ref{fig:s-par}$b$. The use of both CMD and CLF has permitted to derive age estimations in agreement with those from the literature, however a
future extension to the TO region will help in better define the age of some clusters, especially those for which recent deep optical-data have
revealed the presence of multiple populations \citep[e.g. NGC 1806 and NGC 1783,][]{Mackey+08}. Differences between the two calibrations are mostly
due to the efficiency of overshooting in stellar convective H cores \citep[][and references therein]{Barmina+02,Brocato+03}.

\begin{figure*}[t]
 \plottwo{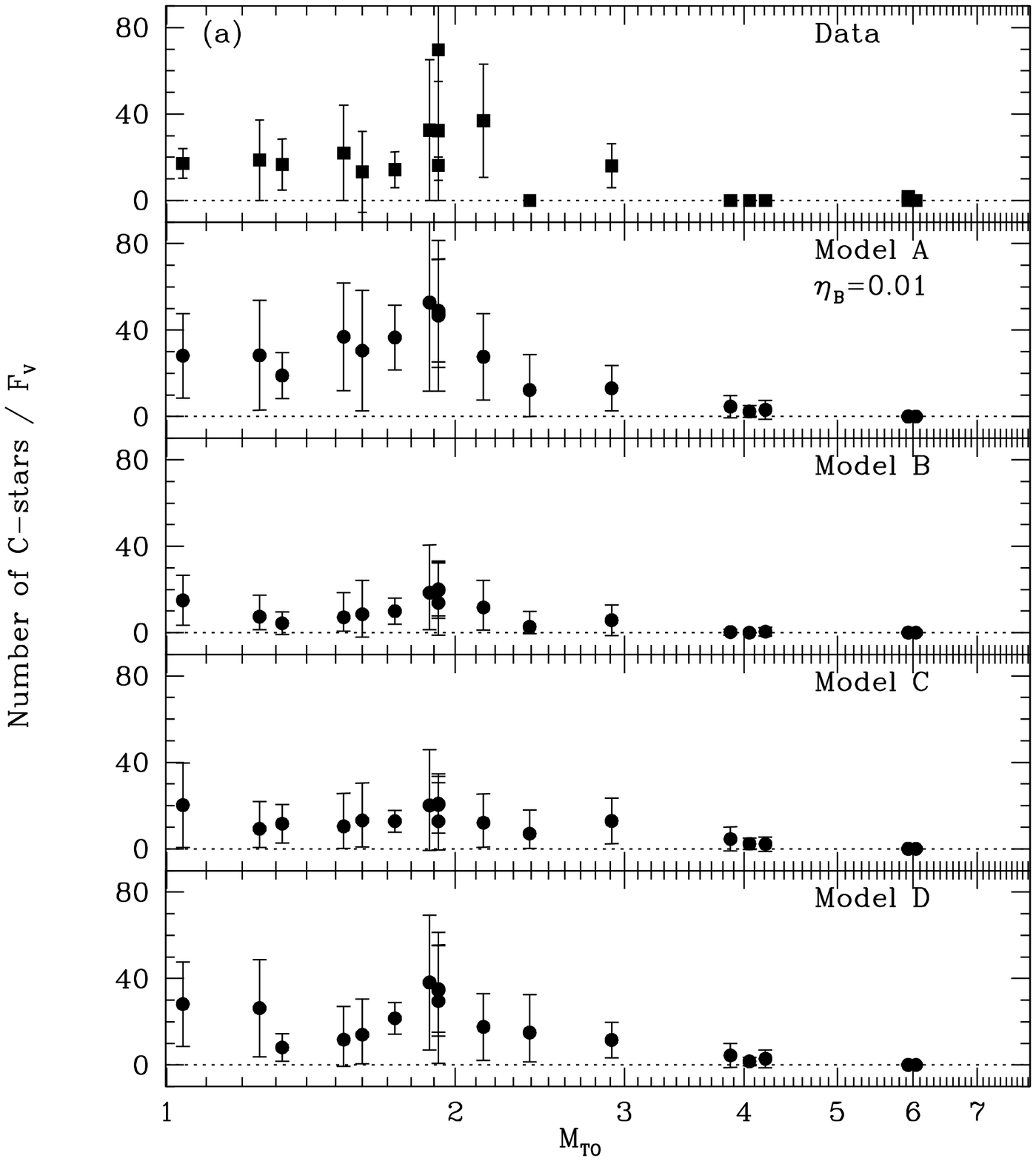}{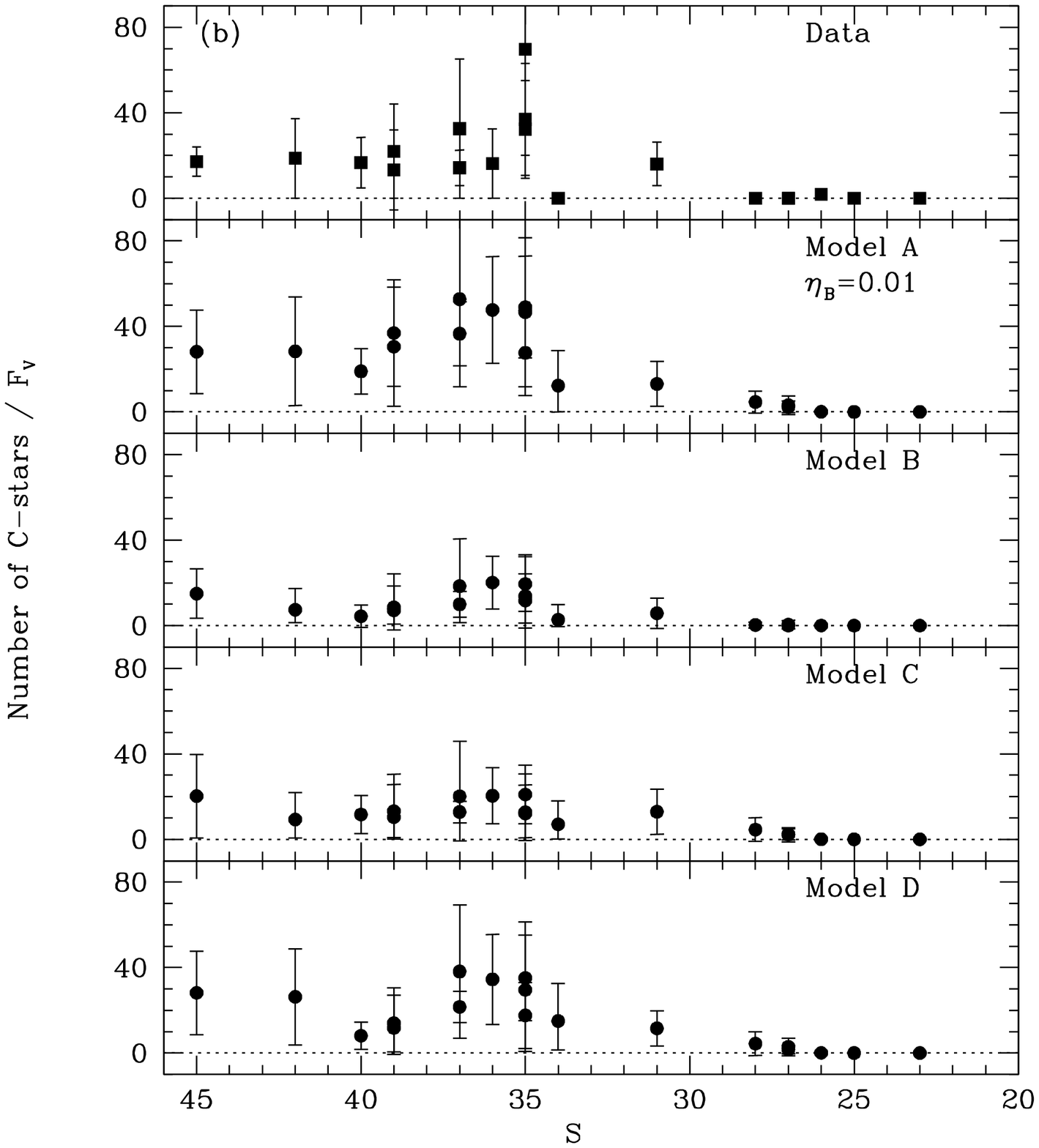}
  \caption{The number of C-rich stars normalized to the total $V$ luminosity of the cluster is reported as a function of the TO mass
  (left) and $s$-parameter (right) for individual clusters. Squares represent data (upper panels), and filled circles synthetic models.}
  \label{fig:MTO}
\end{figure*}

We compare now the expected number of C-rich AGB stars derived from models \emph{A-B-C-D} to the observed one. In Figure~\ref{fig:MTO} the number of
C-rich AGB stars normalized to the total visual light is plotted as a function of the TO mass ($M_{TO}$, Figure~\ref{fig:MTO}$a$) and the
$s$-parameter (Figure~\ref{fig:MTO}$b$). The error budget of data accounts for the uncertainty of integrated visual light and the membership of
C-stars, while theoretical uncertainty is estimated taking into account statistical fluctuations in the number of TP-AGB stars and the uncertainty of
integrated visual light. As shown in Table \ref {tab:cluster_info}, the number and luminosity of bright AGB stars may suffer a large statistical
fluctuation in clusters of similar age. This is due to the small number statistics, as each cluster contains at most a few TP-AGB stars. This is
visible in Figure~\ref{fig:MTO}$a$ at $M_{TO} \sim 1.8-2.2$ M$_\sun$, where a few clusters having nearly the same age posses a different number of
TP-AGB stars. Around these values there is a maximum in the frequency of the TP-AGB stars \citep[see also][]{Marigo&Girardi07}. The qualitative
analysis of Figure~\ref{fig:MTO} also shows that the predicted trend is similar in all models, even thought the observed maximum is nicely reproduced
by \emph{model A}, and to a lesser extent \emph{model D} and $C$. We find that in the mass range 1.7$\lsim$ M/M$_{\sun}$ $\lsim$2.5 and for
metallicity $Z=0.008$ the TP-AGB lifetime is of the order of $\sim$ 2-3 Myr and reduces both at lower and higher masses. This result is in agreement
with the analysis recently performed by \citet{Marigo&Girardi07}.

\subsection{Integrated NIR colors} \label{ss:colors}

TP-AGB stars affect NIR-bands integrated luminosity and its uncertainty in low-metallicity intermediate-age massive clusters \citep[see
e.g.][]{Girardi&Bertelli98, Maraston98, Mouhcine&Lancon02, Maraston05}. In particular, NIR colors may be dominated by a handful of red giant stars
\citep{Santos&Frogel97,Brocato+99a,Cervino+03}. At the typical age of Galactic globular clusters (GGC), the presence of C-rich stars becomes more
uncertain, as in GGC AGB stars are all observed to be O-rich, so that carbon does not appear to have been dredged up into the envelop during thermal
pulses \citep{Lattanzio&Wood03}. It is worth noting that the assumption on mass loss is expected to have a marginal effect on integrated magnitudes
and colors of very faint populations, where bright TP-AGB stars are statistically less frequent, or even absent \citep{Santos&Frogel97,Fagiolini+07}.

\begin{figure*}[t]
  \plotone{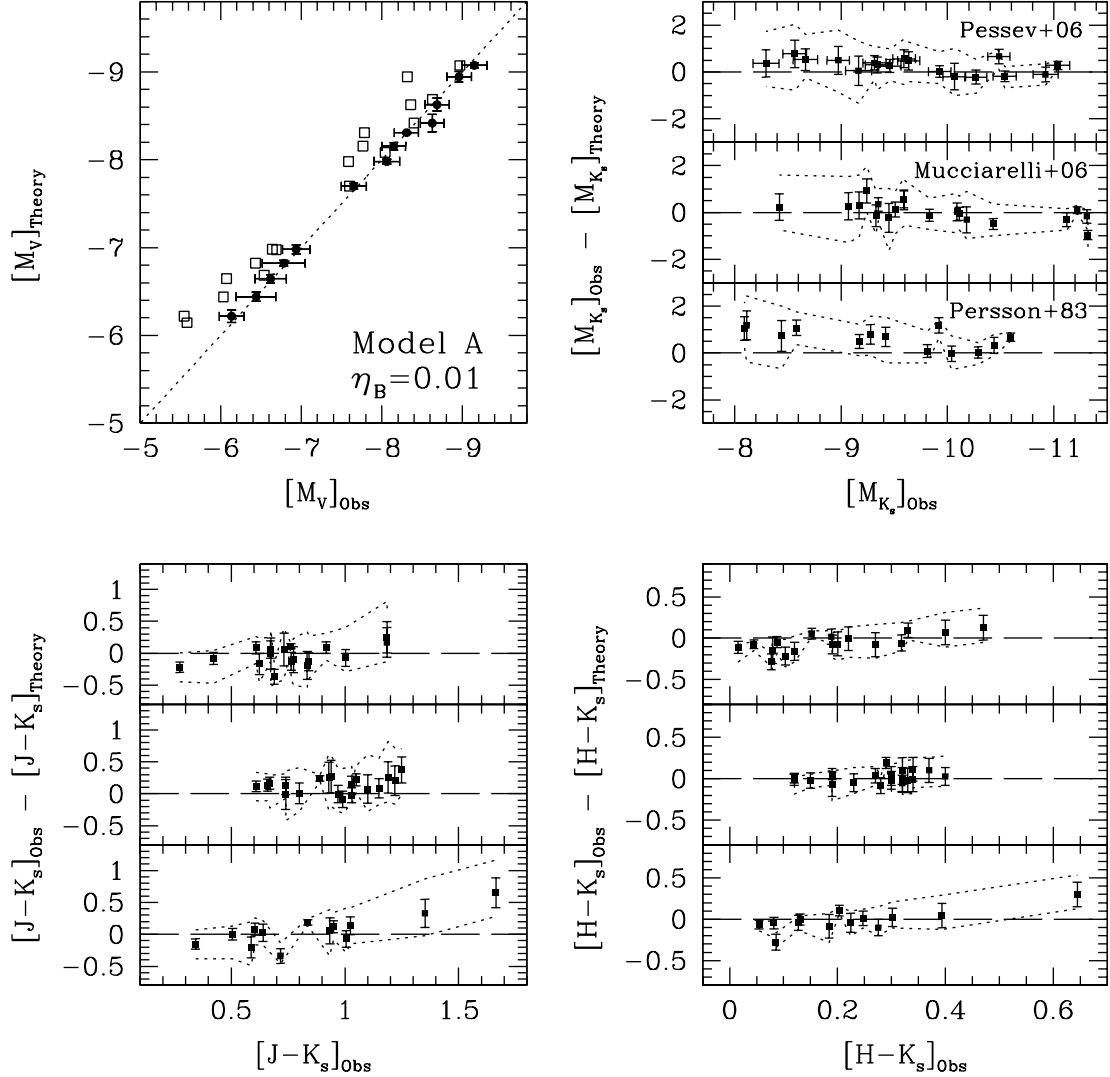}
  \caption{Integrated magnitudes and colors for 19 LMC clusters. Predicted values refer to \emph{Model A}.
  \emph{Upper left panel:} Predicted $M_V$ $vs.$ observed $M_V$ of \citet[][open squares]{VanDenBergh81}, and \citet[][filled circles]{Goudfrooij+06}.
  \emph{Upper right panel:} The $M_{K_s}$ magnitude difference is compared to the observed $M_{K_s}$.
  \emph{Lower panels:} The difference between predicted and observed $J-K_s$ (left) and $H-K_s$ (right) are compared to the observed values
   from three different compilations: \citet{Pessev+06} (upper), \citet{Mucciarelli+06} (middle), and \citet{Persson+83} (bottom).
   The dotted lines mark the region covered by all the available value in the synthetic color distribution, i.e. it is related to the maximum and
   the minimum magnitude (color) obtained in the series of independent runs.}
  \label{fig:colors_bw001}
\end{figure*}

\begin{figure*}[t]
  \plotone{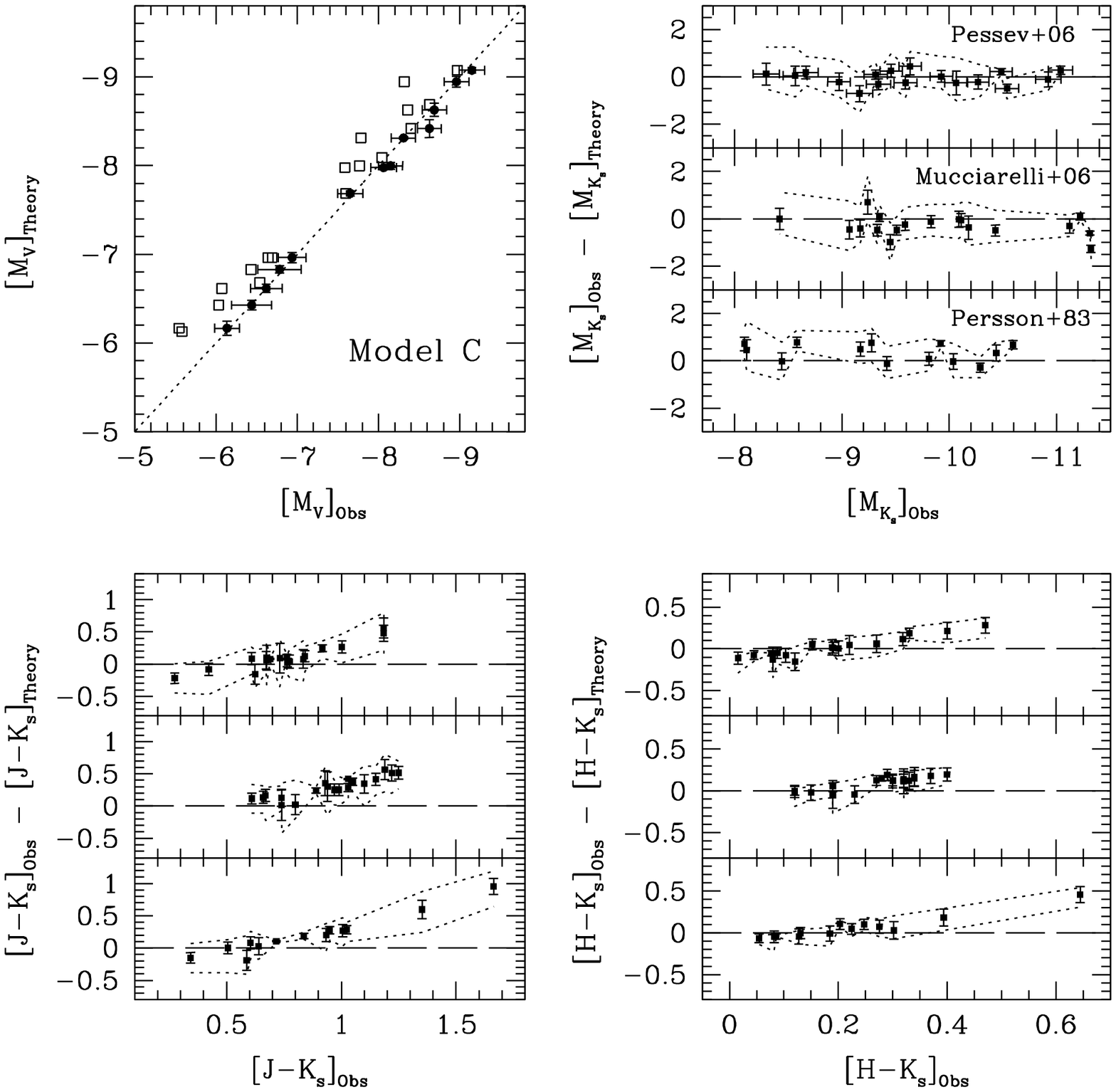}
  \caption{As in Figure \ref {fig:colors_bw001}, but for \emph{Model C}.}
  \label{fig:colors_v05}
\end{figure*}

The differences between observed and predicted magnitudes (colors) of each cluster are plotted for models \emph{A} and \emph{C} in
Figure~\ref{fig:colors_bw001} and Figure \ref{fig:colors_v05}, respectively. Theoretical values are calculated as the average over 200 simulations
(see \S~\ref{ss:imteo}), and error-bars are related to the standard deviation (1$\sigma$). Simulated values of $M_V^{tot}$ are compared to
measurements of \citet{Goudfrooij+06} or \citet{VanDenBergh81} (left higher panel), while predicted values of $M_{K_s}$ and colors ($J-K_s$ and
$H-K_s$) are compared to data from \citet[][top]{Pessev+06}, \citet[][middle]{Mucciarelli+06}, and \citet[][bottom]{Persson+83}. Observations are
decontaminated for the field contribution, and dereddened by using the mean $E_{B-V}$ values labelled in Figure~\ref{fig:cmd}.

Figure~\ref{fig:colors_bw001} shows that the agreement is nice for all the clusters in the case of \citet{Pessev+06} and \citet{Mucciarelli+06},
while a discrepancy can be seen with \citet{Persson+83}. As pointed out by \citet{Pessev+06}, a major discrepancy between their color measurements
with those of \citet{Persson+83} are caused by centering problems and to the presence of several relatively bright stars in the background field. In
the figure it is clearly shown that colors of low-luminosity clusters have a large dispersion \citep[see
also][]{Santos&Frogel97,Brocato+99a,Fagiolini+07}. This is the case of the faint cluster NGC~2209, for which \citet{Pessev+06} recognize the major
difference with \citet{Persson+83}. For this cluster we estimated an age $\log t = 8.95 \pm 0.10$, in agreement with our previous value from
optical-HST data \citep{Raimondo+05a}. The cluster contains two C-stars, identified by \citet{Walker71} and confirmed by \citet{Frogel+90},
dominating NIR fluxes. No M-giants are possessed by the cluster.

\begin{figure*}[t]
  \plotone{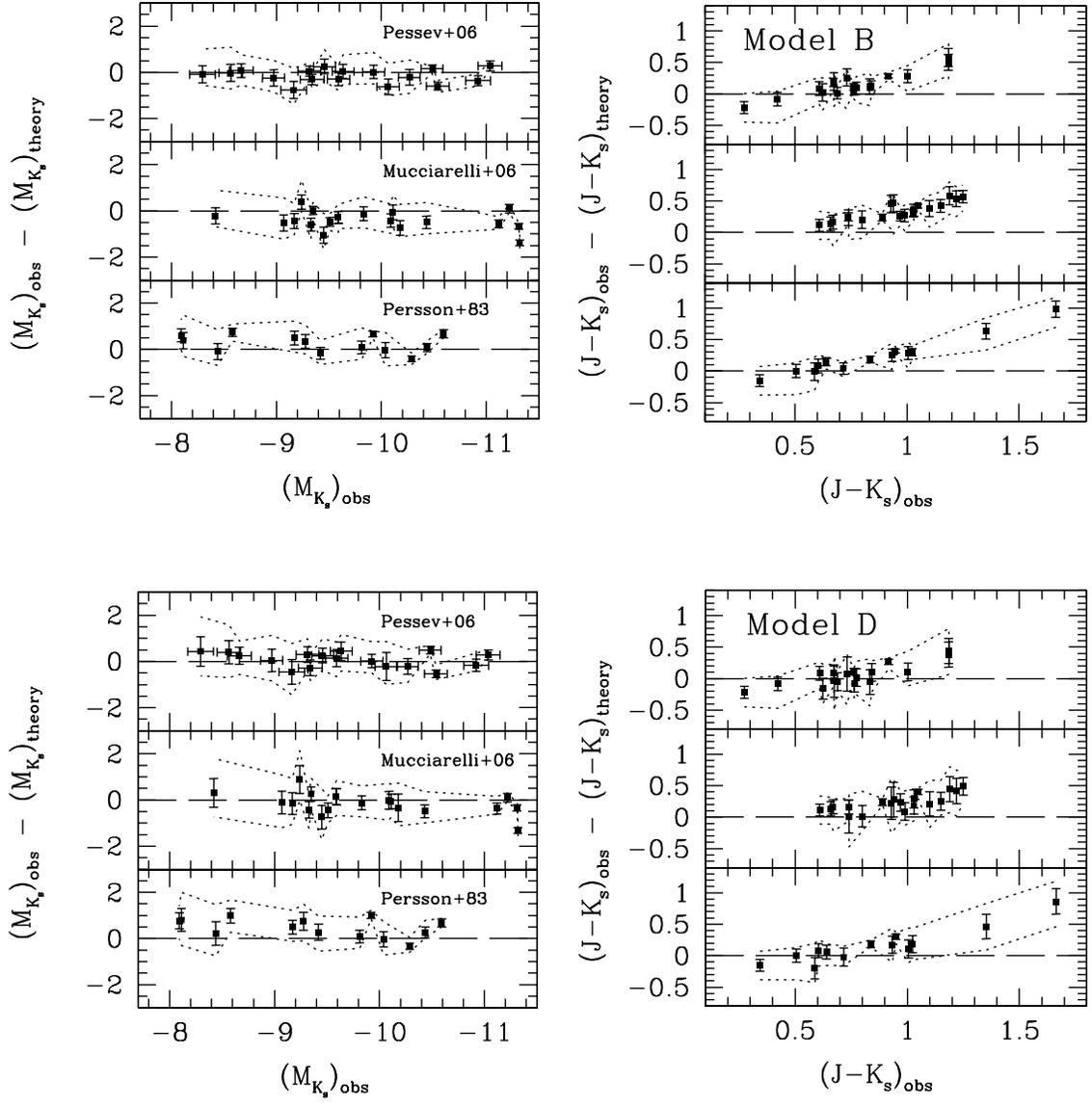}
  \caption{$K_s$-magnitude and $J-K_s$ color differences for model \emph{Model B}
  (upper panels) and \emph{Model D} (lower panel).}
  \label{fig:colors_BD}
\end{figure*}

Inspection in Figure~\ref{fig:colors_v05} shows that predicted $K_s$ magnitudes still agree with observations within the errorbars, especially in the
case of data from \citet{Pessev+06} and \citet{Mucciarelli+06}. With a few exceptions, differences between observed and predicted colors are
contained in the region marked by the dotted lines, illustrating all the color values covered by simulations. Thus, synthetic colors are still able
to reproduce the data in the case of models \emph{C}, even if the agreement worsens at redder colors (i.e. $J-K_s \gsim 1$ mag and $H-K_s \gsim 0.3$
mag), as a consequence of the smaller number of TP-AGB stars predicted by \emph{model C} compared to \emph{model A}. Nearly the same results can be
inferred from the analysis of colors from \emph{model D}, and slightly worsen for \emph{model B} (see Figure~\ref{fig:colors_BD}), where for sake of
brevity only differences between observed and predicted $K_s$-magnitudes and ($J-K_s$) colors are reported.

In conclusion, observed NIR magnitudes and colors of LMC clusters are reproduced by \emph{model A}, and to a lesser extent by \emph{model C},
\emph{D}, and \emph{B} (the worst case), within errorbars or at least within the color distributions obtained from all the simulations (200) computed
for each cluster. Therefore, even though NIR integrated magnitudes and colors are sensitive to TP-AGB stars, they do not appear to be efficient
enough to discriminate among physical assumptions which affect the number of bright TP-AGB stars in clusters, at least for the degree of accuracy of
models proposed here (\S~\ref{ss:TPmassloss}).

\subsection{Surface Brightness Fluctuations} \label{ss:SBF}

In this section we present observed SBF magnitudes of individual clusters and compare them with theoretical predictions. In the case of star clusters
the SBF measurements are obtained from the second moment of the stellar luminosity function directly derived from resolved-star photometry
\citep{Ajhar&Tonry94,Gonzalez+04,Raimondo+05a}. Let us recall that this procedure measures the \emph{same} physical quantity obtained for galaxies,
and derived from the pixel-to-pixel brightness variation in the galaxy image. Studying SBFs of LMC star clusters is then a relevant step to
understand the SBF signal from unresolved galaxies where light is (totally or partially) emitted by the underlying mixture of old and
intermediate-age populations.

SBF magnitudes of clusters were estimated as follows. We coupled individual stellar photometry and total fluxes of the clusters both provided by
\citet{Mucciarelli+06}. The denominator is the cluster total flux, free from field contamination and reddening. The depth of the stellar photometry
used here, contributing to the numerator, is adequate for our purposes. In fact, as discussed by \citet{Ajhar&Tonry94}, in the optical and at least
for halo GGGs, the second moment of the luminosity (numerator), converges quickly, with 99\% of the sum being obtained with the three brightest
magnitudes of cluster stars. This is even more evident in NIR bands \citep{Gonzalez+04}, where the total light is dominated typically by RGB and AGB
stars in populations older than a few hundreds of Myr. For sake of consistency with the analysis in \S~\ref{ss:cmd}, we consider only stars within
1.5$\arcmin$. As already noted, even in this restricted area, field stars are distinguishable from cluster stars only in the youngest clusters, since
in the remaining ones giant stars belonging to the cluster and to the field occupy the same regions of the CMD (see Figure~\ref{fig:cmd}).
Consequently, the contribution of field stars to the numerator can be easily subtracted for the first six clusters. In older clusters we used the
control-field frames, assuming that the field stellar population present in the adjacent field is identical to that populating the cluster's area.
The procedure was applied to each cluster of the sample and results are reported in Table \ref{tab:cluster_sbf}. The errorbars also include the
effect due to the uncertainty on the membership of bright stars.

\begin{figure*}[t]
  \plotone{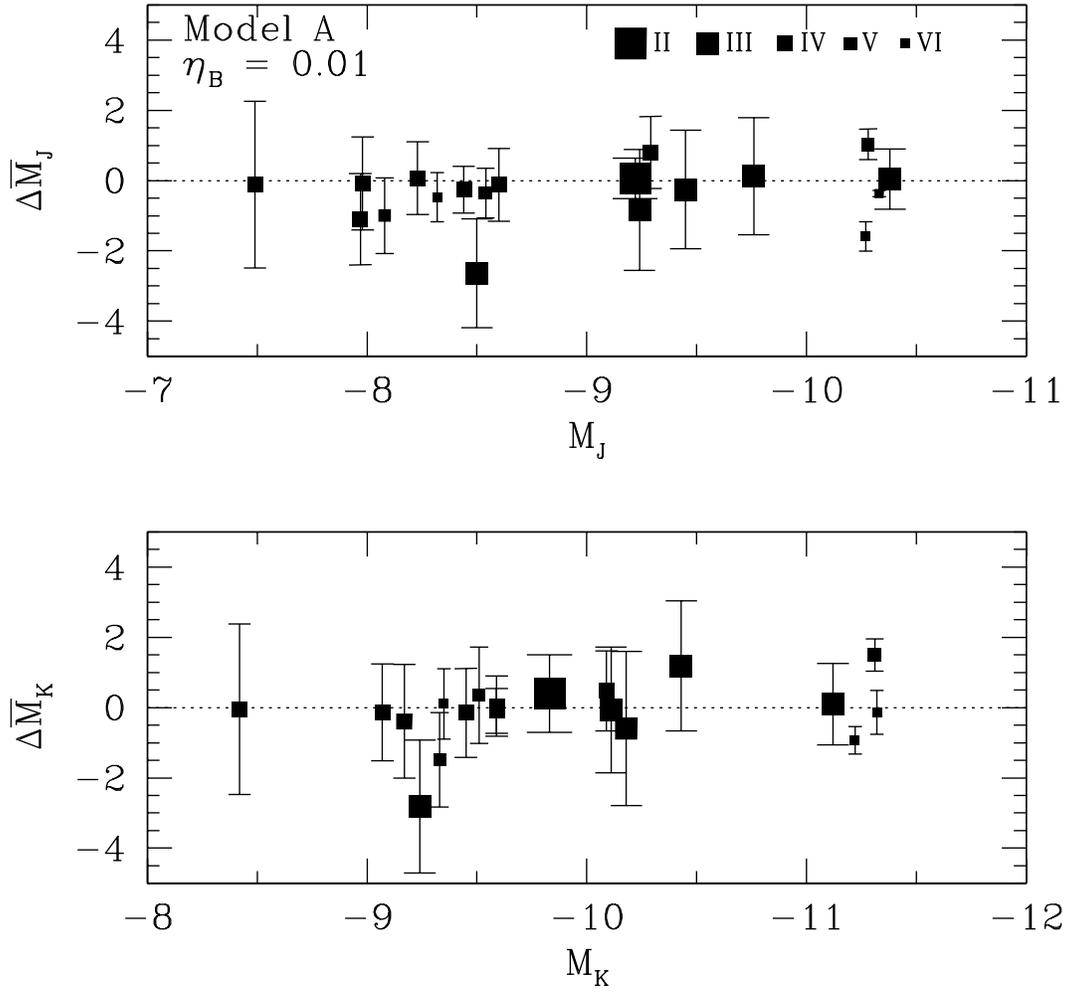}
  \caption{The difference between observed and predicted SBF magnitudes.
  Symbol size decreases from SWB class II to VI as labeled.}
  \label{fig:sbfJK_bw001}
\end{figure*}

\begin{figure*}[t]
  \plotone{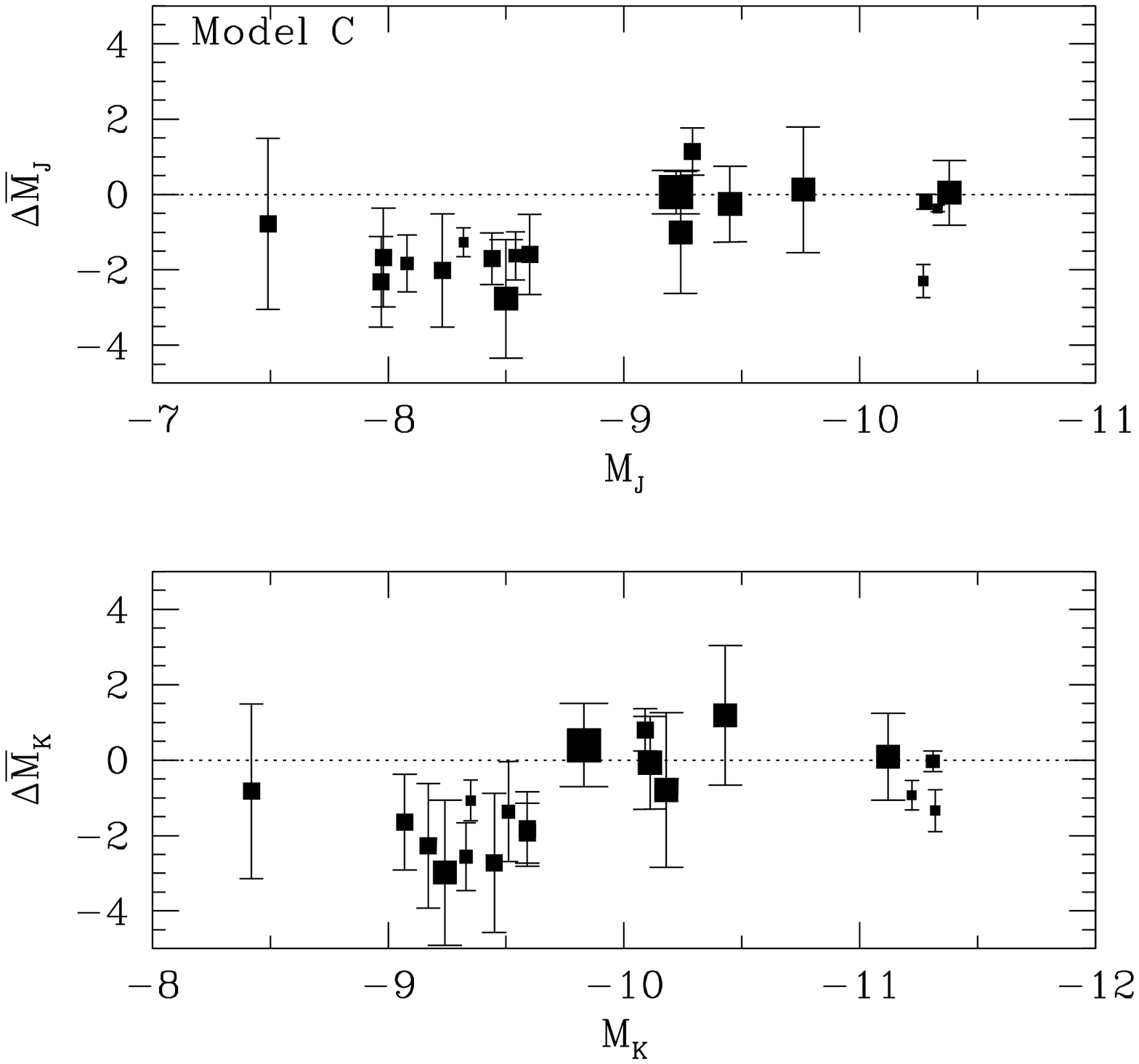}
  \caption{As in Figure \ref {fig:sbfJK_bw001}, but for \emph{Model C}.}
  \label{fig:sbfJK_v05}
\end{figure*}

We calculated theoretical SBFs from the exact simulations used to reproduce the CMDs, CLFs, integrated magnitudes and colors. In practice, for a
given cluster SBF amplitudes were derived applying Eq.~\ref{eq:eqTS} and \ref{eq:eqTS2}. In Figures~\ref{fig:sbfJK_bw001} and \ref{fig:sbfJK_v05} we
plot differences between observed and predicted SBFs (\emph{model A} and \emph{model C}, respectively) as a function of the observed $M_J$ and
$M_{K_s}$ magnitudes. Differently from colors, that are reproduced by both assumptions (even if with different precision), the SBF data/models
comparison shows a clear different behavior. SBFs from \emph{model A} provide a good agreement with data in the entire range of $M_J$ and $M_{K_s}$,
being differences mostly lower or of the order of $\sim 0.5$ mag. There are only three outlayers, namely NGC~2134 and, at less extent, NGC~1783 and
NGC~1806, in which a high field contamination was recognized (Figure~\ref{fig:clf}). On the other hand, \emph{model C} foresees SBFs fainter than the
observed ones, in particular this model underestimates the number of bright stars in the magnitude intervals $-8 \lsim M_J \lsim -9$ and $-9 \lsim
M_K \lsim -9.8$ (Figure~\ref{fig:sbfJK_v05}).

\begin{figure*}[t]
  \plotone{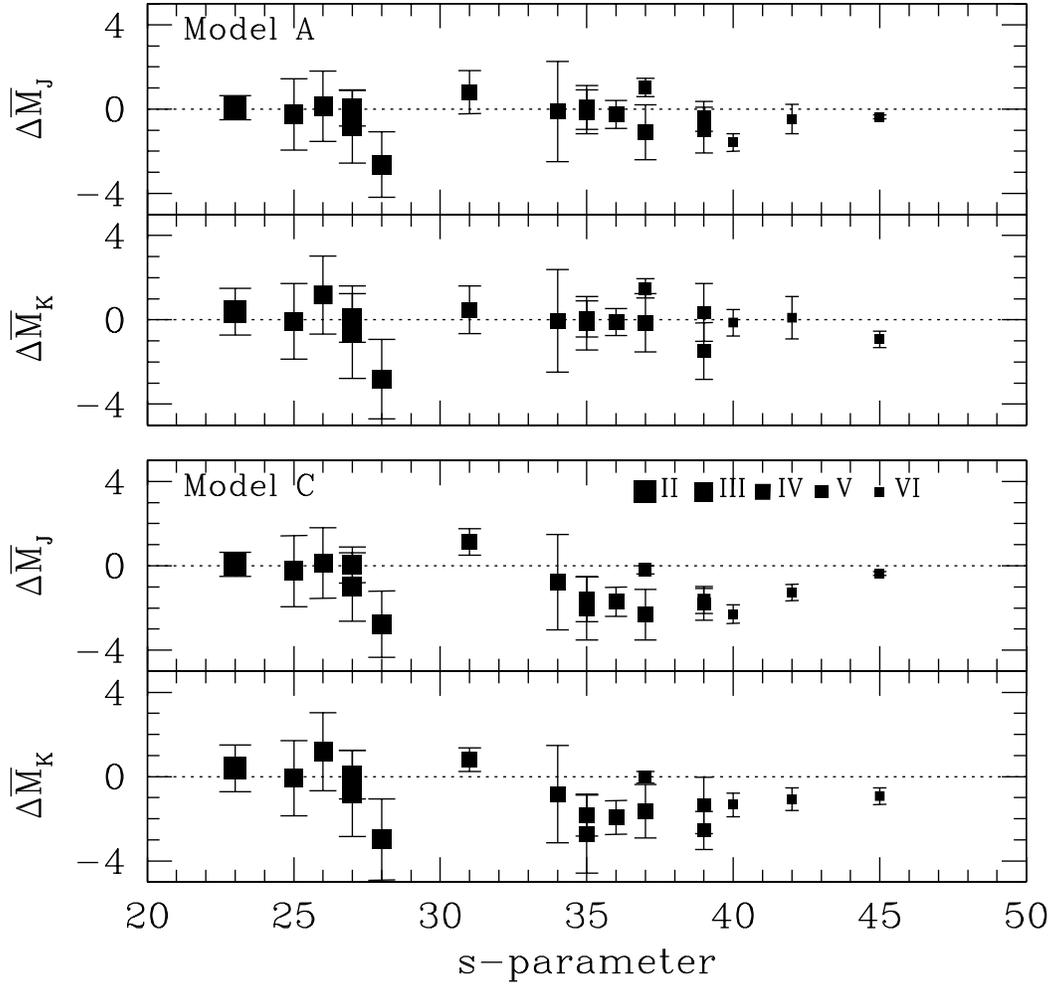}
  \caption{SBF-magnitude differences are plotted against the $s$-parameter.}
  \label{fig:ds}
\end{figure*}

To better interpret these results, Figure~\ref{fig:ds} shows the relationship between SBF differences and s-parameter (age). Again, in the case of
\emph{model A} the agreement is good for all the s-parameter values considered here. On the other hand, it is clear that clusters with $s>30$ (ages
greater than about 200 Myr), i.e. the age at which the number of TP-AGB stars becomes statistically relevant, suffer large discrepancy in the case of
\emph{model C}. Although the sample does not contain very old clusters, our guess is that for ages older than $\sim $5 Gyr the effect should
decrease, as already shown in \cite{Raimondo+05a}. Note that results from \emph{model B} and \emph{D} show a behavior similar to \emph{model C}.
Thus, changing the number of TP-AGB stars mainly affect the SWB classes IV, V and VI (see also Figure~\ref{fig:MTO}). We recall the four models
(\emph{A-D}) differ only for the treatment of the mass-loss process of TP-AGB stars. On this basis, we suggest that \emph{model A} is more
appropriate to reproduce the SBF measurements of LMC star clusters in the range $30 < s < 45$, where TP-AGB stars are expected to have a large impact
on the cluster observational properties. It is worth noting that this conclusion is not unique and should be verified when different assumptions are
used in stellar population models (e.g. IMF, stellar evolutionary tracks, color-temperature relations, TP-AGB synthetic prescriptions, etc.). Within
this limitation, we suggest that SBFs can contribute to disentangle the observable properties of TP-AGB stars in unresolved galaxies containing a
large number of such stars.

\begin{figure*}[t]
  \plotone{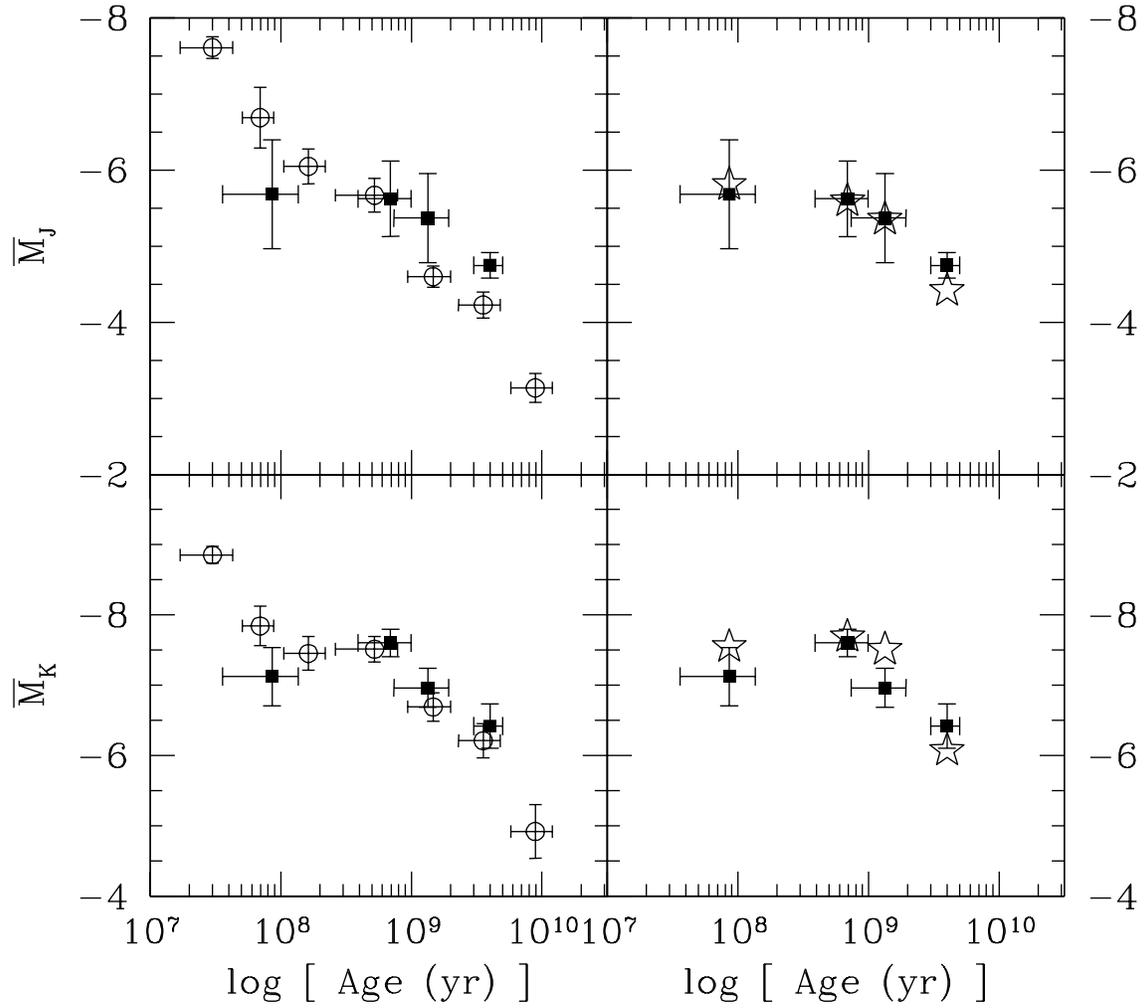}
  \caption{Left Panels: Observed SBF magnitudes of clusters as grouped according to Table \ref{tab:SBF_SWB} (filled squares) are compared to
  Superclusters' data from \citealt{Gonzalez+04}, open circles). Right Panels: Observed SBF magnitudes of clusters as grouped
  according to Table \ref{tab:SBF_SWB} are compare to predictions from \emph{model A} (open stars, see text).}
  \label{fig:JKSWB}
\end{figure*}

Before concluding this section, we compare present SBF measurements to those by \citet{Gonzalez+04}. They derived NIR-SBFs of star clusters in the
MCs as grouped (to obtain a sort of super-clusters) according to SWB classes \citep{Searle+80}. This procedure allows the authors to reduce the
problem of small number statistics, and to compare LMC cluster data to SBF predictions computed for systems with a large number of stars, i.e.
galaxies \citep[see][for details]{Raimondo+05a}. We grouped the present clusters using the same definition of SWB classes. Our sample does not cover
the Pre, SWB-I and VII classes. Moreover, the SWB-II class contains only one cluster (NGC2164), thus it is not significative. Table~\ref{tab:SBF_SWB}
summarizes the cluster grouping and the resulting SBF data. Figure~\ref{fig:JKSWB} compares SBF data of super-clusters from both sets of measures.
Despite the smaller number of clusters considered in the present work, the agreement between the two sets of measurements is quite good, even though
there is a systematic age shift of SWB-III and SWB-V due to the different $s$-parameter/age relation. \citet{Gonzalez+04} assigned ages and
metallicities to the super-clusters from \citet{Frogel+90}, with a few exceptions regarding SWB-I, SWB-II and Pre-SWB super-cluster. The open stars
in Figure~\ref{fig:JKSWB} result from grouping the corresponding synthetic CMDs, i.e. we applied exactly the same procedure used for observations.
The proper SWB class is assigned according to the present age estimations  of each cluster, then a mean is considered. We note a slight discrepancy
($<$0.5 mag) for SWB-VI in the $J$-band and for SWB-V in the $K_s$-band. We recall that for these two classes the overlapping between cluster and
field population is more severe. As a conclusion, we note that the data/models comparison for super-clusters stands again for a consistency of the
present procedure and SPS models.

\section{Summary and conclusions}
\label{s:discussion}

In this paper we carried out a joint analysis of \emph{resolved} and \emph{unresolved} properties of a sample of LMC clusters, in order to
consistently  derive the evolutionary status of stars in the clusters, by using different stellar population synthesis tools based on an unique
theoretical framework. The final aim was understanding the potential of the SBF technique as tracer of stellar populations, and exploring the use of
such a technique to constrain more tightly stellar population models, those of intermediate age in particular.

To reach this goal, we studied the CMD, CLF, integrated magnitudes and colors, and SBFs of 19 LMC star clusters, covering an age range from a few Myr
up to several Gyr. We used accurate NIR ($J$ and $K_s$ bands) photometric data of cluster's resolved stars \citep{Ferraro+04,Mucciarelli+06}, and NIR
colors from up-to-date compilations by \citet{Goudfrooij+06}, \citet{Pessev+06} and \citet{Mucciarelli+06}. By coupling resolved star photometry with
total fluxes, we derived new measurements of NIR-SBFs for each individual cluster, paying attention to field star contamination which may highly
affect stellar counts in the upper part of the luminosity function. This allows us to obtain a set of homogeneous and high-quality SBF data of pure
intermediate-age stellar populations. This is a necessary step toward having a larger sample of star clusters with accurate measurements of SBFs.

We used the multi-purpose synthesis code \emph{SPoT}, based on a unique and consistent theoretical framework, to predict all the quantities quoted
above. Great attention was payed to take into account stochastic effects. This is a crucial point, as broad-band colors and, in general,
spectro-photometric features of clusters may suffer from large \emph{intrinsic} fluctuations caused by the discrete nature of the number of stars
\citep[see e.g.][]{Barbaro&Bertelli77,Chiosi+88,Santos&Frogel97, Brocato+99a, Raimondo+05a}. Therefore, including stochastic effects is an important
step to derive reliable age and metallicity estimations from cluster broad-band colors \citep[e.g.][]{Fagiolini+07}.

We also presented new synthesis models (\emph{A-D} in Sect. \ref{ss:TPmassloss}), computed under the same prescriptions for all the evolutionary
phases except for the mass-loss rates of TP-AGB stars. Model results were compared to data through the fitting of observed CMD, CLF, and integrated
colors of each cluster. The agreement obtained in reproducing both \emph{resolved} and \emph{unresolved} observational quantities suggests a high
degree of reliability of the adopted theoretical framework. The cluster ages and metallicities derived by best-fit models are also in agreement
(within the uncertainties) with previous determinations from other indicators. The age estimates were used to derive a new calibration of the
\emph{s}-parameter/age relationship for intermediate-age stellar clusters, i.e. \emph{s} values within the range 20-45.

Following the same line presented in \citet{Raimondo+05a}, we showed that NIR-SBFs are extremely sensitive to the number and properties of AGB
(especially TP-AGB) stars. Changing the mass-loss rate prescription affects TP-AGB properties, including the duration of the entire phase and thus
SBF predictions change sizably. On the basis of the theoretical framework adopted (WG98 and W90) to describe the TP-AGB evolution, we found that
although all the models are able to reproduce observed CMDs, CLFs and integrated magnitudes and colors, even though with different precision, not all
the models reproduce the SBF measurements of LMC clusters. In particular, only \emph{model A} provides a good fit of NIR-SBF in the age range
corresponding to $30 < s < 45$, where TP-AGB stars are expected to be more numerous. Thus, we argue that the global theoretical scenario adopted in
\emph{model A} appears to be able to reproduce all the observational properties of the LMC star clusters investigated in this work. It is worth
noticing that the theoretical prescriptions used for modeling stellar clusters may affect our results on TP-AGB synthetic evolution, for instance the
luminosity and effective temperature at the first thermal pulse. In this sense, an effort to study the physics of AGB structure experiencing thermal
pulses is urged to provide solid theoretical background for stellar population synthesis studies. In spite of this limit, we showed that NIR-SBFs are
very sensitive to the properties of TP-AGB stars, then they can be useful to investigate the global nature of such stars in clusters and galaxies,
where they are numerous and statistical effects are less important. Using the large and homogeneous sample of clusters presented here, we have shown
that SBF amplitudes may be linked to stellar evolution properties, and, in turn, they can be used as an intriguing and complementary diagnostic to
test the ingredients of SPS models, and evolution properties of bright giant stars.

As conclusive discussion, we notice that the metallicity dependence of the C-rich phase lifetime was accounted for using the relationship between the
number ratio of C- to M-stars and the metallicity observed for the Local Group galaxies \citep{Groenewegen06}. However, the number ratio of
C-rich/O-rich stars should be a complex function of the properties of individual AGB stars, i.e. the durations of the entire phase and the sub-phases
(O/C-rich, respectively), and the location of AGB stars in the CMD. Moreover, even though the use of a purely synthetic code to describe the TP-AGB
phase is less accurate than the full stellar modeling, it is the only way to explore a large parameter space in computing SPS models. The
uncertainties in assigning --as well as in predicting-- stellar parameters of variable TP-AGB stars (e.g. effective temperature, opacity,
luminosity), where effects such as convection, variability, mass loss, and dust formation become increasingly important \citep[][]{Freytag&Hofner08},
still have a great impact in SPS models. Modern stellar evolutionary codes as well as hydrodynamical codes for stellar atmospheres are at present
extensively used to model different aspects of stellar physics at work in stars beyond the core-helium burning phase, as the interior nucleosynthesis
and envelope chemical enrichment and their dependence on stellar metallicity and mass, hence we expect to reduce the uncertainty of the predicted
properties of AGB stars and then of intermediate-age stellar population models in the next future. Once again detailed analysis and accurate
simulations of well-known stellar populations like LMC star clusters represent a fundamental step to improve our understanding of the nature of
contributors to the integrated light from remote galaxies, as well as the potential of SBFs to disentangle the evolutionary status of unresolved
populations.

\begin{acknowledgements}

We warmly thank Alessio Mucciarelli and Francesco Ferraro for providing us photometric data, and Adriano Pietrinferni and Santi Cassisi for updated
stellar evolution models. It is a pleasure to thank Enzo Brocato and Michele Cantiello for useful discussions on the subject. This work received
financial support by INAF-PRIN/06 (PI G.C.). We acknowledge the anonymous referee for comments and suggestions which greatly improved the paper.

\end{acknowledgements}



\clearpage


\clearpage

\begin{deluxetable}{lccccccc}
\tabletypesize{\scriptsize} \tablecaption{LMC Cluster Sample Information\label{tab:cluster_info}} \tablewidth{0pt} \tablehead{ \colhead{Cluster Name}
& \colhead{Log Age} & \colhead{Reference} & \colhead{[Fe/H]} & \colhead{Reference} & \colhead{C-stars } & \colhead{Reference}
& \colhead{$s$} \\
    & (dex) &     & (dex) &     & (Number) &     &    \\
(1) & (2)   & (3) & (4)   & (5) & (6)      & (7) & (8) } \startdata
NGC 1651 & 9.30$^{+0.08}_{-0.10}$ & 5   & $-$0.37$\pm$0.20& 7      & 1;1;1 & 1;9;11  & 39\\
         & 9.08                   & 1   & $-$0.53$\pm$0.03& 12     &       &         &   \\
NGC 1783 & 8.94                   & 1   & $-$0.45         & 8      & 2;4;4 & 1;9;11  & 37\\
NGC 1806 & 9.16                   & 1   & $-$0.23         & 7      & 4;2   & 1;9     & 40\\
NGC 1831 & 8.50$\pm$0.30          & 4   &    0.01$\pm$0.20& 7      & 3;2   & 1;11    & 31\\
         & 8.58                   & 1   &                 &        &       &         &   \\
NGC 1866 & 8.12$\pm$0.30          & 4   & $-$0.50$\pm$0.10& 6      & 0;0   & 1;9     & 27\\
         & 8.20                   & 1   &                 &        &       &         &   \\
NGC 1978 & 9.52                   & 1   & $-$0.96         & 6      & 4;4;9 & 1;9;11  & 45\\
NGC 1987 &  8.79                  & 1   & $-$1.00         & 8      & 3;1;1 & 1;9;11  & 35\\
NGC 2031 & 8.20$\pm$0.10          & 2   & $-$0.52$\pm$0.21& 2      & 0     & 1       & 27\\
         & 8.20                   & 1   &                 &        &       &         &   \\
NGC 2108 & 8.86                   & 1   & $-$1.20         & 8      & 1;1;1 & 1;9;11  & 36\\
NGC 2134 & 8.28                   & 1   & $-$1.00         & 8      & 0;0   & 1;11    & 28\\
NGC 2136 & 8.00$\pm$0.10          & 2   & $-$0.55$\pm$0.23& 2      & 1;0;0 & 1;9;11  & 26\\
         & 8.13                   & 1   &                 &        &       &         &   \\
NGC 2157 & 7.60$\pm$0.20          & 3   & $-$0.45         & 10     & 0     & 1       & 25 \\
         & 8.06                   & 1   & $-$0.60         & 8      &       &         &   \\
NGC 2162 & 9.11$^{+0.12}_{-0.16}$ & 5   & $-$0.23$\pm$0.20& 7      & 1;0   & 1;11    & 39 \\
         & 9.08                   & 1   & $-$0.46$\pm$0.07& 12     &       &         &   \\
NGC 2164 & 7.70$\pm$0.20          & 4   & $-$0.45         & 10     & 0     & 1       & 23\\
         & 7.91                   & 1   & $-$0.60         & 8      &       &         &   \\
NGC 2173 & 9.33$^{+0.07}_{-0.09}$ & 6   & $-$0.24$\pm$0.20& 7      & 1;1;1 & 1;9;11  & 42\\
         & 9.30                   & 1   & $-$0.42$\pm$0.03& 12     &       &         &   \\
NGC 2190 & 8.86                   & 1   & $-$0.12         & 7      & 2;2   & 1;11    & - \\
NGC 2209 & 8.98$^{+0.15}_{-0.24}$ & 5   & $-$0.47         & 10     & 2;2;2 & 1;9;11  & 35\\
         & 8.79                   & 1   & $-$1.20         & 8      &       &         &   \\
NGC 2231 & 9.18$^{+0.10}_{-0.13}$ & 5   & $-$0.67$\pm$0.20& 7      & 1;1;1 & 1;9;11  & 37\\
         & 8.94                   & 1   & $-$0.52$\pm$0.03& 12     &       &         &   \\
NGC 2249 & 8.82$\pm$0.30          & 4   & $-$0.47         & 10     & 0     & 1       & 34\\
         & 8.72                   & 1   & $-$0.12         & 8      &       &         &   \\
\enddata
\tablecomments{ Col. (1): Cluster's name. Col. (2): Cluster's Age from literature. Col. (3) Reference of data in Col. (2). Col. (4): Cluster's
metallicity from the literature. Col. (5): Reference of data in Col. (4). Col. (6): Number of detected C-stars. Col. (7): Reference of data in Col.
(6). Col. (8): $s$ parameter.} \tablecomments{References: (1) \citet{Mucciarelli+06}, (2) \citet{Dirsch+00}, (3) \citet{Elson91}, (4)
\citet{Elson&Fall88}, (5) \citet{Geisler+97}, (6) \citet{Hill+00}, (7) \citet{Olszewski+91}, (8) \citet{Sagar&Pandey89}, (9) \citet{Frogel+90}, (10)
\citet{Mackey&Gilmore03}, (11) \citet{Westerlund+91}, (12) \citet{Grocholski+06}.}
\end{deluxetable}


\begin{deluxetable}{ccccccc}
\tabletypesize{\scriptsize} \tablecaption{Synthetic properties of cluster sample\label{tab:cluster_synthetic}} \tablewidth{0pt} \tablehead{
\colhead{Cluster} & \colhead{log Age} & \colhead{Z} & \colhead{$M_V$} & \colhead{$M_{K_s}$} & \colhead{$J-K_S$} & \colhead{$H-K_s$} \\
 Name             & (yr)         &  (dex)  & (mag)          & (mag)          & (mag)         & (mag) \\
(1)          & (2)            & (3)      & (4)              & (5)              & (6)           & (7) } \startdata

NGC 2164     & 7.70$^{+0.18}_{-0.15}$ &  2.0e-02 & $-$8.42$\pm$0.10 & $-$9.7$\pm$0.3   & 0.49$\pm$0.09 & 0.12$\pm$0.05 \\
NGC 2157     & 7.72$^{+0.17}_{-0.15}$ &  2.0e-02 & $-$8.68$\pm$0.06 &$-$10.0$\pm$0.3   & 0.52$\pm$0.11 & 0.13$\pm$0.07 \\
NGC 2136     & 7.72$^{+0.17}_{-0.37}$ &  2.0e-02 & $-$8.62$\pm$0.09 & $-$9.9$\pm$0.3   & 0.50$\pm$0.11 & 0.14$\pm$0.07 \\
NGC 1866     & 8.08$^{+0.07}_{-0.10}$ &  8.0e-03 & $-$9.07$\pm$0.03 &$-$10.8$\pm$0.3   & 0.61$\pm$0.15 & 0.17$\pm$0.09 \\
NGC 2031     & 8.08$\pm$0.08          &  2.0e-02 & $-$8.07$\pm$0.05 & $-$9.9$\pm$0.6   & 0.67$\pm$0.28 & 0.22$\pm$0.16 \\
NGC 2134     & 8.18$^{+0.05}_{-0.10}$ &  2.0e-02 & $-$7.99$\pm$0.04 &$-$10.2$\pm$0.6   & 0.75$\pm$0.25 & 0.25$\pm$0.15 \\
NGC 1831     & 8.50$\pm$0.07          &  2.0e-02 & $-$8.07$\pm$0.02 &$-$10.1$\pm$0.4   & 0.79$\pm$0.18 & 0.28$\pm$0.11 \\
NGC 2249     & 8.68$\pm$0.09          &  8.0e-03 & $-$6.69$\pm$0.06 & $-$8.7$\pm$0.6   & 0.68$\pm$0.26 & 0.22$\pm$0.14 \\
NGC 2190     & 8.81$\pm$0.07          &  8.0e-03 & $-$6.73$\pm$0.06 & $-$9.5$\pm$0.6   & 0.93$\pm$0.26 & 0.33$\pm$0.15 \\
NGC 2209     & 8.95$\pm$0.10          &  8.0e-03 & $-$6.33$\pm$0.07 & $-$9.2$\pm$0.7   & 1.01$\pm$0.25 & 0.34$\pm$0.15 \\
NGC 1987     & 8.95$\pm$0.10          &  8.0e-03 & $-$7.04$\pm$0.05 &$-$10.1$\pm$0.4   & 1.08$\pm$0.14 & 0.37$\pm$0.10 \\
NGC 2108     & 8.95$\pm$0.10          &  8.0e-03 & $-$7.28$\pm$0.05 &$-$10.1$\pm$0.5   & 1.06$\pm$0.17 & 0.37$\pm$0.11 \\
NGC 2231     & 8.98$\pm$0.11          &  8.0e-03 & $-$6.29$\pm$0.08 & $-$9.3$\pm$0.6   & 1.02$\pm$0.24 & 0.34$\pm$0.15 \\
NGC 1783     & 9.08$^{+0.15}_{-0.23}$ &  8.0e-03 & $-$8.28$\pm$0.04 &$-$11.1$\pm$0.3   & 1.05$\pm$0.13 & 0.36$\pm$0.10 \\
NGC 2162     & 9.18$\pm$0.06          &  8.0e-03 & $-$6.44$\pm$0.06 & $-$9.2$\pm$0.5   & 0.88$\pm$0.22 & 0.27$\pm$0.14 \\
NGC 1651     & 9.23$^{+0.07}_{-0.12}$ &  8.0e-03 & $-$6.60$\pm$0.06 & $-$9.6$\pm$0.4   & 0.97$\pm$0.15 & 0.32$\pm$0.11 \\
NGC 1806     & 9.34$^{+0.10}_{-0.14}$ &  4.0e-03 & $-$7.72$\pm$0.03 &$-$10.4$\pm$0.2   & 0.83$\pm$0.11 & 0.23$\pm$0.09 \\
NGC 2173     & 9.48$\pm$0.08          &  8.0e-03 & $-$6.89$\pm$0.03 & $-$9.7$\pm$0.3   & 0.89$\pm$0.15 & 0.26$\pm$0.11 \\
NGC 1978     & 9.70$\pm$0.08          &  4.0e-03 & $-$8.95$\pm$0.01 &$-$11.3$\pm$0.2   & 0.65$\pm$0.05 & 0.10$\pm$0.07 \\
\enddata
\end{deluxetable}


\begin{deluxetable}{ccccc}
\tabletypesize{\scriptsize} \tablecaption{SBF magnitudes of LMC clusters \label{tab:cluster_sbf}} \tablewidth{0pt} \tablehead{ &
               \multicolumn{2}{c}{Models} &            \multicolumn{2}{c}{Observations}
\\
\colhead{Cluster Name} &
\colhead{$\overline M_J$} & \colhead{$\overline M_{K_s}$}& \colhead{$\overline M_J$}& \colhead{$\overline M_{K_s}$}\\
                                   & (mag)                     & (mag)            &(mag)  &  (mag) \\
 (1)         & (2)                 & (3)              & (4)              & (5)   } \startdata
 NGC 2164    & $-5.1 \pm   0.6$    & $-6.1 \pm 0.8$ & $-5.1 \pm 0.2$  & $-5.7 \pm 0.8$ \\
 NGC 2157    & $-5.3 \pm   1.0$    & $-6.4 \pm 1.3$ & $-5.6 \pm 1.4$  & $-6.5 \pm 1.3$ \\
 NGC 2136    & $-5.2 \pm   0.7$    & $-6.3 \pm 1.0$ & $-5.0 \pm 1.5$  & $-5.1 \pm 1.6$ \\
 NGC 1866    & $-5.5 \pm   0.9$    & $-7.0 \pm 1.2$ & $-5.5 \pm 0.1$  & $-7.0 \pm 0.1$ \\
 NGC 2031    & $-5.5 \pm   1.8$    & $-7.0 \pm 2.2$ & $-6.4 \pm 0.1$  & $-7.6 \pm 0.3$ \\
 NGC 2134    & $-5.8 \pm   1.6$    & $-7.4 \pm 1.9$ & $-8.5 \pm 0.1$  & $-10.3 \pm 0.1$ \\
 NGC 1831    & $-5.5 \pm   1.0$    & $-7.4 \pm 1.2$ & $-4.8 \pm 0.3$  & $-6.9 \pm 0.1$ \\
 NGC 2249    & $-4.5 \pm   1.8$    & $-5.9 \pm 2.2$ & $-4.6 \pm 1.5$  & $-6.0 \pm 1.0$ \\
 NGC 2190    & $-5.7 \pm   1.0$    & $-7.5 \pm 1.3$ & $-5.7 \pm 0.1$  & $-7.8 \pm 0.1$ \\
 NGC 2209    & $-5.6 \pm   1.4$    & $-7.4 \pm 1.7$ & $-5.6 \pm 0.2$  & $-7.6 \pm 0.1$ \\
 NGC 1987    & $-5.9 \pm   0.5$    & $-7.9 \pm 0.6$ & $-6.1 \pm 0.9$  & $-7.8 \pm 0.6$ \\
 NGC 2108    & $-5.9 \pm   0.6$    & $-7.8 \pm 0.7$ & $-6.2 \pm 0.4$  & $-8.0 \pm 0.1$ \\
 NGC 2231    & $-5.6 \pm   1.0$    & $-7.4 \pm 1.3$ & $-6.7 \pm 0.8$  & $-7.5 \pm 0.2$ \\
 NGC 1783    & $-5.6 \pm   0.5$    & $-7.7 \pm 0.5$ & $-4.7 \pm 0.1$  & $-6.3 \pm 0.1$ \\
 NGC 2162    & $-5.0 \pm   1.0$    & $-6.7 \pm 1.4$ & $-6.0 \pm 0.4$  & $-8.3 \pm 0.6$ \\
 NGC 1651    & $-5.3 \pm   0.7$    & $-7.2 \pm 0.9$ & $-5.7 \pm 0.3$  & $-6.9 \pm 1.1$ \\
 NGC 1806    & $-5.0 \pm   0.5$    & $-6.9 \pm 0.6$ & $-6.9 \pm 0.1$  & $-7.1 \pm 0.1$ \\
 NGC 2173    & $-4.7 \pm   0.7$    & $-6.6 \pm 1.0$ & $-5.2 \pm 0.1$  & $-6.5 \pm 0.1$ \\
 NGC 1978    & $-4.3 \pm   0.1$    & $-5.4 \pm 0.1$ & $-4.7 \pm 0.1$  & $-6.4 \pm 0.4$ \\
\enddata
\end{deluxetable}


\begin{deluxetable}{lllcccc}
\tabletypesize{\scriptsize} \tablecaption{Cluster grouping and SBF magnitudes\label{tab:SBF_SWB}} \tablewidth{0pt} \tablehead{ & & &
\multicolumn{2}{c}{Models} &  \multicolumn{2}{c}{Observations} \\
\colhead{SWB Class} & \colhead{$s$} &\colhead{ Cluster Name } &\colhead{$\overline M_J$} &\colhead{$\overline M_{K_s}$} &\colhead{$\overline M_J$} &
\colhead{ $\overline M_{K_s}$ } } \startdata

II    & 22-24 & 2164                           & \nodata & \nodata  & \nodata          & \nodata \\
III   & 25-29 & 2157,2136,1866,2031,2134       & $-5.8$  & $-7.5$   & $-5.7 \pm 0.7$ & $-7.1 \pm 0.4$ \\
IV    & 30-36 & 1831,2249,1987,2209,2190,2108  & $-5.6$  & $-7.7$   & $-5.6 \pm 0.5$ & $-7.6 \pm 0.2$ \\
V     & 37-41 & 2231,1806,1651,2162,1783       & $-5.4$  & $-7.5$   & $-5.4 \pm 0.6$ & $-7.0 \pm 0.3$ \\
VI    & 42-46 & 2173,1978                      & $-4.4$  & $-6.1$   & $-4.7 \pm 0.2$ & $-6.4 \pm 0.3$ \\
\enddata
\end{deluxetable}

\end{document}